\documentclass[prd,preprint,tightenlines,floatfix,showpacs,preprintnumbers,nofootinbib,eqsecnum,cernpreprint]{revtex4-1}

\usepackage{amsfonts}
\usepackage{amsmath}
\usepackage{hyperref}
\usepackage{amssymb}
\usepackage[english]{babel}
\usepackage{graphicx}
\usepackage{epsfig}
\usepackage{bm}
\usepackage{verbatim}
\usepackage[utf8]{inputenc}
\usepackage{color}
\usepackage[dvipsnames]{xcolor}
\usepackage[normalem]{ulem}
\usepackage{hyperref}
\usepackage{subcaption}
\usepackage{physics}

\usepackage{feynmp-auto}

\usepackage{slashed}

\usepackage{multirow}  
\usepackage{makecell} 
\usepackage{booktabs}

\usepackage{tikz}
\usetikzlibrary{shapes.geometric,arrows}

\newcommand{\mathsym}[1]{{}}

\input{tcilatex}

	\allowdisplaybreaks
	
\begin{document}
		
	\title{Gravitational waves from a scotogenic two-loop neutrino mass model}
	
	\begin{flushright}
	 	CERN-TH-2023-071\\
		\vskip1cm
	\end{flushright}
	
	\author{Cesar Bonilla$^{1}$}
	\email{cesar.bonilla@ucn.cl}
	
	\author{A. E. C\'{a}rcamo Hern\'{a}ndez$^{2,3,4}$}
	\email{antonio.carcamo@usm.cl}
		
	\author{Jo\~{a}o Gonçalves$^{5,6}$}
	\email{jpedropino@ua.pt}
	
	\author{Vishnudath K. N.$^{2}$}
	\email{vishnudath.neelakand@usm.cl}

	\author{Antonio P.~Morais$^{5,7}$}
	\email{aapmorais@ua.pt}

	\author{R. Pasechnik$^{6}$}
	\email{roman.pasechnik@thep.lu.se}

	\affiliation{$^1$\sl Departamento de Física, Universidad Católica del Norte, Avenida Angamos 0610, Casilla 1280, Antofagasta, Chile
	\\
	{$^2$\sl Universidad T\'ecnica Federico Santa Mar\'{\i}a, Casilla 110-V, Valpara\'{\i}so, Chile}
	\\
	{$^3$\sl Centro Cient\'{\i}fico-Tecnol\'ogico de Valpara\'{\i}so, Casilla 110-V,Valpara\'{\i}so, Chile}
	\\
	{$^4$\sl Millennium Institute for Subatomic Physics at High-Energy Frontier (SAPHIR), Fern\'andez Concha 700, Santiago, Chile}
	\\
	{$^5$\sl Departamento de F\'{\i}sica da Universidade de Aveiro and Centre for \\ Research and Development in Mathematics and Applications (CIDMA), \\ Campus de Santiago, 3810-183 Aveiro, Portugal}
	\\
	{$^6$\sl Department of Physics, Lund University, 221 00 Lund, Sweden}
	\\
	{$^7$\sl Theoretical Physics Department, CERN, 1211 Geneva 23, Switzerland}	
	}
	
	\begin{abstract}
		We propose a framework to account for neutrino masses at the two-loop level. This mechanism introduces new scalars and Majorana fermions to the Standard Model. It is assumed the existence of a global $\mathrm{U(1)\times \mathcal{Z}_2}$ symmetry which after partial breaking provides the stability of the dark matter candidates of the theory.
		The rich structure of the potential allows for the possibility of first-order phase transitions (FOPTs) in the early Universe which can lead to the generation of primordial gravitational waves. Taking into account relevant constraints from lepton flavour violation, neutrino physics as well as the trilinear Higgs couplings at next-to-leading order accuracy, we have found a wide range of possible FOPTs which are strong enough to be probed at the proposed gravitational-wave interferometer experiments such as LISA.
	\end{abstract}
	
	\maketitle
	\tableofcontents
		
	\flushbottom
	\newpage
	
	\section{Introduction}\label{sec:intro}
	
	The Standard Model (SM) is a highly successful theory that describes the electromagnetic, strong and weak interactions whose predictions have been experimentally verified at the LHC with the highest degree of accuracy. However it has several unaddressed issues such, for example the current pattern of SM fermion masses and mixing angles, the number of SM fermion families, the measured amount of dark matter relic density and baryon asymmmetry observed in the Universe, among others. Experiments with solar, atmospheric, and reactor neutrinos have brought evidence of neutrino oscillations caused by nonzero masses. Several extensions of the SM have been constructed in order to explain the tiny masses of the active neutrinos; see \textit{e.g.} Ref.~\cite{Cai:2017jrq} for a review and a non-extensive list \cite{Jana:2019mgj,Arbelaez:2022ejo,Rosenlyst:2021tdr,Cacciapaglia:2020psm} of some comprehensive studies of one and two loop radiative neutrino mass models. The most economical way to generate the tiny masses of the active neutrinos considering the SM gauge symmetry, is by adding two right handed Majorana neutrinos that mix with the light active neutrinos thus triggering a type I seesaw mechanism ~\cite{Minkowski:1977sc,Yanagida:1979as,Glashow:1979nm,Mohapatra:1979ia,Gell-Mann:1979vob, Schechter:1980gr, Schechter:1981cv}, where either the right handed Majorana neutrinos have to be extremelly heavy with masses of the order of the Grand Unification scale or they can be around the TeV scale thus implying very tiny Dirac Yukawa couplings, in order to successfully reproduce the neutrino data. In both scenarios, the mixing between the active and sterile neutrinos is very tiny thus leading to strongly suppressed charged lepton flavor (clfv) violating signatures, several orders of magnitude below the experimental sensitvity, thus making this scenario untestable via clfv decays. This makes models with tree-level type-I seesaw realizations difficult to test via charged-lepton flavor-violating decays. Alternatively, radiative seesaw models are examples of interesting and testable extensions of the SM explaining tiny neutrino masses. In most radiative seesaw models, the tiny neutrino masses arise at a one-loop level, thus implying that in order to successfully reproduce the experimental neutrino data, one has to rely either on very small neutrino Yukawa couplings (of the order of the electron Yukawa coupling) or on an unnaturally small mass splitting between the CP-even and CP-odd components of the neutral scalar mediators. Two-loop neutrino mass models have been proposed in the literature~\cite{Bonilla:2016diq,Baek:2017qos,Saad:2019vjo,Nomura:2019yft,Arbelaez:2019wyz,Saad:2020ihm,Xing:2020ezi,Chen:2020ptg,Nomura:2020dzw} to provide a more natural explanation for the tiny active neutrino masses than those ones based on one-loop level radiative seesaw mechanisms. In this work we propose a minimally extended IDM theory where the scalar sector is enlarged by the inclusion of two electrically neutral gauge singlet scalars and the fermion sector is augmented by adding 4 right handed Majorana neutrinos. The gauge symmetry of the SM model is extended by including a spontaneously broken $U(1)_X$ global lepton number symmetry as well as a preserved $Z_2$ discrete symmetry. The scalar sector of our model is similar to the one of Ref. \cite{Kajiyama:2013rla} however the fermion sector is more economical since in our model there are four right handed Majorana neutrinos whereas the model of \cite{Kajiyama:2013rla} includes nine right handed Majorana neutrinos in the fermionic spectrum. Additionally the model  of Ref. \cite{Kajiyama:2013rla} has a $U(1)_{B-L}$ gauge symmetry which is not present in our model, thus leading to a different phenomenology. On the other hand, unlike the model of Ref. \cite{Nomura:2020dzw}, our model does not rely in doubly charged scalar fields to implement the two loop level realization of active neutrino masses.
	
	\section{Theoretical structure of the model}\label{sec:model}
	
	\subsection{Particle content and charge assignments}
	
	In this work, we augment the SM gauge groups with an extended global abelian symmetry $\mathrm{U(1)} \times \mathcal{Z}_2$, where the $\mathrm{U(1)}$ associated with the global lepton number is spontaneously broken while the $\mathcal{Z}_2$ is preserved. Besides the standard particle content of the SM, additional fields are present in the spectrum, including an inert SU(2) doublet scalar field, $\eta$, and two scalar singlets, $\varphi$ and $\sigma$. In our setup, $\sigma$ acquires a nonzero vacuum expectation value (VEV) while $\varphi$ does not, such that the potential DM candidates can arise from the neutral components of $\eta$ or from $\varphi$. Additional Majorana fermions are also included into this scheme, namely, two $N_R$ fields and two $\Psi_R$. While the electroweak (EW) gauge group does not distinguish $N_R$ and $\Psi_R$, the extended symmetry group treats both differently. 
	
\begin{figure}[htb!]
		\centering
		\includegraphics[width=0.35\textwidth]{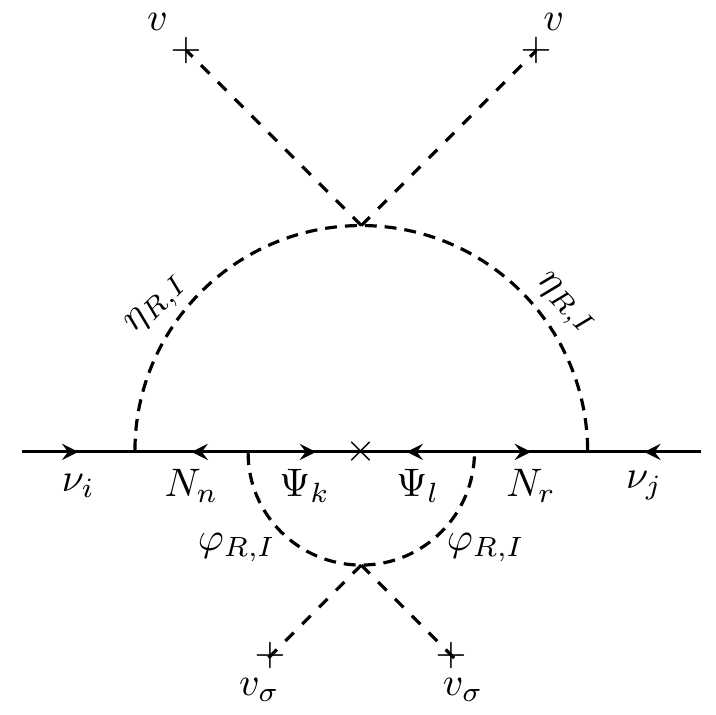}
		\caption{Two-loop Feynman diagram which is responsible for the generation of the neutrino masses and mixing. Here, $\nu$ are the neutrinos, $\eta_{R,I}$ are the real and imaginary parts of the neutral component of the $\eta$ doublet, $\varphi_{R,I}$ are the real and imaginary parts of the $\varphi$ singlet, $v$ and $v_\sigma$  are the VEVs of the $\Phi$ doublet the $\sigma$ singlet fields, respectively.}
		\label{fig:Neutrinoloopdiagram}
	\end{figure}

	The charge assignments of the fields are shown in Tab.~\ref{tab:charges_scalars} for the scalar fields and in Tab.~\ref{tab:charges_fermions} for matter fields. As shown in Tab.~\ref{tab:charges_scalars}, the SU(2) scalar doublet $\eta$ and the scalar singlet $\varphi$ have non-trivial $\mathcal{Z}_2$ charges. Since these two scalars do no acquire VEVs and the charge assignments do not let neutrinos to couple to the SM Higgs, light active neutrinos do not acquire masses at tree level. The specific $\mathrm{U(1)} \times \mathcal{Z}_2$ charge assignments of the $N_{k,\mathrm{R}}$ Majorana neutrinos and $\sigma$ scalar field enables the radiative generation of neutrino mass and the corresponding mixing only at two-loop level that can be noticed from a typical topology in Fig.~\ref{fig:Neutrinoloopdiagram}. Namely, this is possible through the virtual exchange of the neutral components of the inert SU(2) scalar doublet $\eta$ fields as well as of the real and imaginary parts of the inert singlet scalar $\varphi$, with the loop closing through the $\sigma$ VEV and the VEV of the active Higgs doublet $\Phi$. For simplicity of the analysis, we assume that the charged lepton sector is purely diagonal, such that the only contribution to the Pontecorvo–Maki–Nakagawa–Sakata (PMNS) mixing matrix comes exclusively from the loop topology shown in Fig.~\ref{fig:Neutrinoloopdiagram}.
	\begin{table}[htb!]
		\centering
		\begin{tabular}{|cccccc|}
			\hline
			& $\mathrm{SU(3)_C}$ & $\mathrm{SU(2)_L}$ & $\mathrm{U(1)_Y}$  & $\mathrm{U(1)_X}$  & $\mathcal{Z}_2$    \\[0.2em] \hline
			$\Phi$               & $\bm{1}$           & $\bm{2}$           & $1/2$              & $0$                & $1$               \\[0.1em]
			$\sigma$             & $\bm{1}$           & $\bm{1}$           & $0$                & $-1$               & $1$              \\[0.1em]
			$\eta$               & $\bm{1}$           & $\bm{2}$           & $1/2$              & $0$                & $-1$               \\[0.1em]
			$\varphi$            & $\bm{1}$           & $\bm{1}$           & $0$                & $-1$             & $-1$              \\[0.1em] \hline
		\end{tabular}
		\caption{Charge assignments under the SM gauge group ($\mathrm{SU(3)_C} \times \mathrm{SU(2)_L} \times \mathrm{U(1)_Y}$) and the supplemental global symmetry $\mathrm{U(1)_X} \times \mathcal{Z}_2$ for the new scalar fields ($\sigma$, $\eta$ and $\varphi$) and the Higgs doublet ($\Phi$).}
		\label{tab:charges_scalars}
	\end{table}
	\begin{table}[htb!]
		\centering
		\captionsetup{justification=raggedright}
		\begin{tabular}{|cccccc|}
			\hline
			& $\mathrm{SU(3)_C}$ & $\mathrm{SU(2)_L}$ & $\mathrm{U(1)_Y}$  & $\mathrm{U(1)_X}$  & $\mathcal{Z}_2$    \\[0.2em] \hline
			$\ell_{i,\mathrm{L}}$         & $\bm{1}$           & $\bm{2}$           & $-1/2$             & $1$              & $1$               \\[0.1em]
			$\ell_{i,\mathrm{R}}$         & $\bm{1}$           & $\bm{1}$           & $-1$               & $1$              & $1$               \\[0.1em]
			$N_{k,\mathrm{R}}$            & $\bm{1}$           & $\bm{1}$           & $0$                & $1$              & $-1$              \\[0.1em]
			$\Psi_{k,\mathrm{R}}$         & $\bm{1}$           & $\bm{1}$           & $0$                & $0$                & $1$              \\[0.1em]\hline
		\end{tabular}
		\caption{Charge assignments under the SM gauge group ($\mathrm{SU(3)_C} \times \mathrm{SU(2)_L} \times \mathrm{U(1)_Y}$) and the supplemental global symmetry $\mathrm{U(1)_X} \times \mathcal{Z}_2$ for the new matter fields ($N_R$, $\Psi_R$) and the SM-like lepton doublet ($\ell_{i,\mathrm{L}}$) and singlet ($\ell_{i,\mathrm{R}}$). Here, $i=1,2,3$ and $k=1,2$.}
		\label{tab:charges_fermions}
	\end{table}
	
	\subsection{Yukawa sector and scalar potential}
	
	Based on the charge assignments of Tabs.~\ref{tab:charges_scalars} and \ref{tab:charges_fermions}, the most general and renormalisable Yukawa Lagrangian reads as 
	\begin{equation}\label{eq:Yukawa_lag}
		\begin{aligned}
			-\mathcal{L}_{Y}^{(l)} = &(y_{\ell})_{ij}~\overline{\ell}_{i,\mathrm{L}} \Phi \ell_{j,\mathrm{R}} +(y_{N})_{ik}~\overline{\ell}_{i,\mathrm{L}} \tilde{\eta} N_{k,\mathrm{R}} + (y_{\Omega})_{nk}~\overline{N}_{n,\mathrm{R}} \varphi^{\ast} \Psi_{k,\mathrm{R}}^{c} + (m_{\Psi})_{nm}~\overline{\Psi}_{n,\mathrm{R}} \Psi_{m,\mathrm{R}}^{c} + \mathrm{h.c.}\,,
		\end{aligned}  
	\end{equation}
	where we have defined $\tilde{\eta} = i\sigma_2\eta^\dagger$ and the superscript $c$ in the $\Psi$ field indicates charge conjugation. We adopt the Einstein summation convention where repeated indices indicate sum over them, with $i,j=1,2,3$ and $m,n,k = 1,2$. Here, $y_\ell$ is a $3\times 3$ matrix, $y_\Omega$ and $m_\psi$ are $2\times 2$ matrices and $y_N$ is a $2\times 3$ matrix. The neutrino mass matrix can then be induced by the combination of the $y_N$ and $y_\Omega$ Yukawa matrices.
	
	The complete gauge invariant scalar potential in this model is given as,
	\begin{equation}\label{eq:scalar_potential}
		\begin{aligned}
			V=& -\mu _{\Phi }^{2}(\Phi ^{\dagger }\Phi ) + \mu _{\eta }^{2}(\eta ^{\dagger}\eta ) + \mu _{\varphi }^{2}(\varphi ^{\ast }\varphi )-\mu _{\sigma}^{2}(\sigma ^{\ast }\sigma ) 
			 +\lambda_{1}(\Phi ^{\dagger }\Phi )(\Phi ^{\dagger }\Phi )+\lambda_{2}(\eta ^{\dagger }\eta )(\eta ^{\dagger }\eta )  \notag \\
			& +\lambda _{3}(\varphi ^{\ast }\varphi )(\varphi ^{\ast }\varphi )+\lambda_{4}(\sigma ^{\ast }\sigma )(\sigma ^{\ast }\sigma )+\lambda _{5}(\Phi^{\dagger }\Phi )(\eta ^{\dagger }\eta )
			+\lambda _{6}(\Phi ^{\dagger }\eta)(\eta ^{\dagger }\Phi )+\frac{\lambda _{7}}{2}\left[ (\Phi ^{\dagger }\eta)^{2} + \mathrm{h.c.} \right]   \notag \\
			& +\lambda _{8}(\Phi ^{\dagger }\Phi )(\sigma ^{\ast }\sigma )+\lambda_{9}(\Phi ^{\dagger }\Phi )(\varphi ^{\ast }\varphi )+\lambda _{10}(\eta^{\dagger }\eta )(\sigma ^{\ast }\sigma )
			+\lambda _{11}(\eta ^{\dagger }\eta)(\varphi ^{\ast }\varphi )+\lambda _{12}(\varphi ^{\ast }\varphi )(\sigma^{\ast }\sigma )  \notag \\
			& +\frac{\lambda _{13}}{2}\left[ \varphi ^{2}\left( \sigma ^{\ast }\right)^{2} + \mathrm{h.c.} \right]+\lambda_{14}\left[\left(\Phi^{\dagger }\eta\right)\left(\varphi\sigma^{\ast }\right)+ \mathrm{h.c.}\right]+\lambda_{15}\left[\left(\Phi^{\dagger }\eta\right)\left(\sigma\varphi^{\ast }\right)+ \mathrm{h.c.}\right] \\
  			& +\left[ -\mu _{sb}^{2} \sigma ^{2}+\kappa_1 \sigma^3 + \kappa_2 \sigma^2 \sigma^* + \kappa_3 (\Phi^\dag \Phi )\sigma + \kappa_4 (\eta^\dag \eta) \sigma + \kappa_5 (\varphi^*\varphi) \sigma + \mathrm{h.c.} \right], 
		\end{aligned}
	\end{equation}
	where we take all parameters of the potential to be real, i.e.~CP symmetry is conserved in the considered scenario. Notice that the quartic couplings $\lambda_7$ and $\lambda_{13}$ are crucial for generating neutrino masses through out the two-loop level mechanism. The nonzero scalar VEVs are denoted as $\langle \Phi \rangle = v \sim 246~\mathrm{GeV}$ and $\langle \sigma \rangle = v_\sigma$. Also, the terms in the last line of the potential that breaks the $\mathrm{U(1)_X}$ symmetry softly are incorporated in order to make the CP-odd scalar associated with $\sigma$ massive. To simplify our analysis, we take the couplings $\kappa_i$ to be extremely small fom here onwards. Once the scalars $\Phi$ and $\sigma$ acquire VEVs and the EW symmetry is broken, the model will have nine physical scalars -- two charged scalars $(\eta^{\pm})$, four neutral CP-even scalars ($\eta_{R}, \, \varphi_{R},  \,h_1 $ and $h_2$) and three neutral CP-odd scalars ($\eta_{I}, \, \varphi_{I}$ and $\sigma_I$). It is to be noted that the scalars associated with $\eta$ and $\varphi$ mix only among themselves and so do the scalar fields coming from  $\Phi$ and $\sigma$\footnote{Let us note that similar scalar potentials can be found in scenarios that explain Dirac neutrino masses at the one-loop level ~\cite{Bonilla:2018ynb}. A detailed  investigation of such scenarios, following the lines of the present work, will be dedicated in a subsequent paper.}. 
	
	\subsection{Scalar mass spectrum}
	
	After employing the minimization conditions along $v$ and $v_\sigma$, the scalar mass matrices can be written as
	\begin{equation}\label{eq:mass_matrices}
		\begin{aligned}
			&M^2_{\eta^\pm} = \begin{pmatrix}
				0 & 0 & 0 & 0 \\
				0 & 0 & 0 & 0 \\
				0 & 0 & \mu_\eta^2 + \dfrac{v_\sigma^2 \lambda_{10} + v^2\lambda_5}{2} & 0\\
				0 & 0 & 0 & \mu_\eta^2 + \dfrac{v_\sigma^2 \lambda_{10} + v^2\lambda_5}{2}
			\end{pmatrix} \,, \\[0.5em]
			&M^2_{\mathrm{CP-even}} = \begin{pmatrix}
				2v^2\lambda_1 & v v_\sigma \lambda_8 & 0 & 0 \\
				v v_\sigma \lambda_8 & 2 v_\sigma^2 \lambda_4 & 0 & 0 \\
				0 & 0 & \mu_\eta^2 + \dfrac{v_\sigma^2\lambda_{10} + v^2\lambda_5 + v^2\lambda_6 + \lambda_7 v^2}{2} & {\dfrac{1}{4} v v_\sigma \qty(\lambda_{14} + \lambda_{15})} \\
				0 & 0 & {\dfrac{1}{4} v v_\sigma \qty(\lambda_{14} + \lambda_{15})} & \mu_\varphi^2 + \dfrac{v_\sigma^2 \lambda_{12} + v_\sigma^2 \lambda_{13} + v^2\lambda_9}{2}
			\end{pmatrix} \,, \\[0.5em]
			&M^2_{\mathrm{CP-odd}} = \begin{pmatrix}
				0 & 0 & 0 & 0 \\
				0 & 4\mu_{sb}^2 & 0 & 0 \\
				0 & 0 & \mu^2_\eta + \dfrac{ v_\sigma^2\lambda_{10} + v^2\lambda_5 + v^2\lambda_6 - v^2\lambda_7 }{2} & {\dfrac{1}{4} v v_\sigma \qty(\lambda_{15} - \lambda_{14})} \\
				0 & 0 & {\dfrac{1}{4} v v_\sigma \qty(\lambda_{15} - \lambda_{14})} & \mu_\varphi^2 + \dfrac{v_\sigma^2 \lambda_{12} - v_\sigma^2\lambda_{13} + v^2\lambda_9 }{2}
			\end{pmatrix} \,, 
		\end{aligned}
	\end{equation}
	where the zero-eigenvalues of the charged and CP-odd matrices are absorbed by the longitudinal components of the $W^\pm$ and $\mathrm{Z^0}$ gauge bosons. To avoid constraints coming from the Higgs sector, we work closely to the alignment limit, such that $vv_\sigma \lambda_8 \ll 1$. The above matrices can be analytically diagonalised, which results in the mass eigenvalues,
	\begin{equation}\label{eq:mass_eigenvalues}
		\begin{aligned}
			&\mathrm{Charged~scalars} :
			\begin{cases}
				m_{\eta^{\pm}}^2 = \mu_\eta^2 + \dfrac{1}{2} \left(\lambda_5 v^2 +\lambda_{10} v_\sigma^2\right) \,.
			\end{cases} 
			\\
			&\mathrm{CP-even~scalars} :
			\begin{cases}
				& m_{h_1}^{2} = \lambda_1 v^2 + \lambda_4 v_\sigma^2 - \sqrt{(\lambda_1 v^2 - \lambda_4 v_\sigma^2)^2  + \lambda_8^2 v^2 v_\sigma^2   } \,, \\[0.6em]
				& m_{h_2}^{2} = \lambda_1 v^2 + \lambda_4 v_\sigma^2 + \sqrt{(\lambda_1 v^2 - \lambda_4 v_\sigma^2)^2  + \lambda_8^2 v^2 v_\sigma^2   } \,, \\[0.6em]
				& {m_{\eta_{R}}^{2} =  \frac{1}{4} \Big[ 2 (\mu^2_\eta + \mu^2_\varphi )+\text{$\lambda^H_+$} v^2 - \Big\{\Big(2 (\mu^2_\eta -\mu^2_\varphi )+\text{$\lambda^H_-$} v^2\Big)^2 +} \\ &\hspace{4.5em} {2 \text{$v_\sigma $}^2 \Big[2(\mu^2_\eta -\mu^2_\varphi ) (\text{$\lambda_{10}$}-\text{$\lambda_{12}$}-\text{$\lambda_{13}$}) + }\\ &\hspace{4.5em} {v^2 \{\text{$\lambda^H_-$} (\text{$\lambda_{10}$}-\text{$\lambda_{12}$}-\text{$\lambda_{13}$})+2 (\text{$\lambda_{14}$}+\text{$\lambda_{15}$})^2\}\Big]+\text{$\Lambda^H_-$}^2 \text{$v_\sigma $}^4\Big\}^{1/2} + \text{$\Lambda^H_+$}\text{$v_\sigma$}^2 \Big] \,,}\\[0.6em]	  
				& {m_{\varphi_{R}}^{2} =  \frac{1}{4} \Big[ 2 (\mu^2_\eta + \mu^2_\varphi )+\text{$\lambda^H_+$} v^2 - \Big\{\Big(2 (\mu^2_\eta -\mu^2_\varphi )+\text{$\lambda^H_-$} v^2\Big)^2 +} \\ &\hspace{4.5em} {2 \text{$v_\sigma $}^2 \Big[2(\mu^2_\eta -\mu^2_\varphi ) (\text{$\lambda_{10}$}-\text{$\lambda_{12}$}-\text{$\lambda_{13}$}) + }\\ &\hspace{4.5em} {v^2 \{\text{$\lambda^H_-$} (\text{$\lambda_{10}$}-\text{$\lambda_{12}$}-\text{$\lambda_{13}$})+2 (\text{$\lambda_{14}$}+\text{$\lambda_{15}$})^2\}\Big]+\text{$\Lambda^H_-$}^2 \text{$v_\sigma $}^4\Big\}^{1/2} + \text{$\Lambda^H_+$}\text{$v_\sigma$}^2 \Big] \,.}\\[0.6em] 
			\end{cases}
			\\
			&\mathrm{CP-odd~scalars} :
			\begin{cases}
				& m_{\sigma_I}^{2} = 4 \mu _{sb}^{2} \,, \\[0.6em]
				& {m_{\eta_{I}}^{2} =  \frac{1}{4} \Big[ 2 (\mu^2_\eta +\mu^2_\varphi )+\text{$\lambda_+^A$}v^2-\Big\{\Big(2 (\mu^2_\eta -\mu^2_\varphi )+\text{$\lambda^A_-$}v^2\Big)^2 + }\\ &\hspace{4.5em} {2 \text{$v_\sigma $}^2 (2 \text{$\Lambda^A_-$} (\mu^2_\eta-\mu^2_\varphi )+v^2 \{2 (\text{$\lambda_{14}$}-\text{$\lambda_{15}$})^2+\text{$\lambda^A_-$} \text{$\Lambda^A_-$} \} ) +} \\&\hspace{4.5em} {\text{$\Lambda^A_-$}^2 \text{$v_\sigma $}^4\Big\}^{1/2}+\text{$\Lambda^A_+$} \text{$v_\sigma $}^2 \Big]\,,} \\[0.6em]
				& {m_{\varphi_{I}}^{2} =  \frac{1}{4} \Big[ 2 (\mu^2_\eta +\mu^2_\varphi )+\text{$\lambda_+^A$}v^2-\Big\{\Big(2 (\mu^2_\eta -\mu^2_\varphi )+\text{$\lambda^A_-$}v^2\Big)^2 + }\\ &\hspace{4.5em} {2 \text{$v_\sigma $}^2 (2 \text{$\Lambda^A_-$} (\mu^2_\eta-\mu^2_\varphi )+v^2 \{2 (\text{$\lambda_{14}$}-\text{$\lambda_{15}$})^2+\text{$\lambda^A_-$} \text{$\Lambda^A_-$} \} ) +} \\&\hspace{4.5em} {\text{$\Lambda^A_-$}^2 \text{$v_\sigma $}^4\Big\}^{1/2}+\text{$\Lambda^A_+$} \text{$v_\sigma $}^2 \Big] \,.} 
			\end{cases}  
		\end{aligned}
	\end{equation}
	Here, the lightest scalar $m_{h_1}$ is taken to be the SM Higgs boson of mass $\sim 125$ GeV. {To simplify the above expressions, we have defined $\lambda^H_{\pm} = \lambda_5 + \lambda_6 + \lambda_7 \pm \lambda_9$, $\Lambda^H_\pm = \pm \lambda_{10} + \lambda_{12} + \lambda_{13}$ in the CP-even sector and $\lambda^A_\pm = \lambda_5 + \lambda_6 - \lambda_7 \pm \lambda_9$, $\Lambda^A_\pm = \lambda_{10} \pm \lambda_{12} \mp \lambda_{13}$ in the CP-odd sector. To make the numerical analysis simpler, we work in the limit of small mixings between the two heavy CP-even and between the two CP-odd states (effectively we take $\lambda_{14}$ and $\lambda_{15}$ to be small). Under this assumption,} the above expressions can be inverted to express 7 out of the 13 quartic couplings in terms of the physical scalar masses. Thus, the physical masses can be taken as the input parameters in the numerical analysis. The relations that are used in this work are,
	\begin{equation}\label{eqn:inverted_eqs} 
		\begin{aligned}
			\lambda_1 &~= \frac{m^2_{h_1}-\sqrt{(m^4_{h_1}-2 m^2_{h_2} m^2_{h_2} +m^4_{h_2}-4 \lambda_8^2v_1^2 v_\sigma^2 )}+m^2_{h_2} }{4v_1^2} \,, \\ 
			\lambda_4 &~= \frac{m^2_{h_1}+\sqrt{ \left(m^4_{h_1}-2 m^2_{h_1} m^2_{h_2}+m^4_{h_2}-4 \lambda_8^2 v_1^2 v_\sigma^2\right)}+m^2_{h_2}}{4 v_\sigma^2} \,, \\
			\lambda_5 &~{\approx} \frac{-2 \mu_\eta ^2+2 m_{\eta ^{\pm }}^2-\lambda_{10} v_\sigma^2}{v_1^2} \,, \\
			\lambda_6 &~{\approx} \frac{m_{\eta_{I}}^{2} + m_{\eta_{R}}^{2} - 2 m_{\eta ^{\pm }}^{2} }{v_1^2} \,, \\
			\lambda_7 &~{\approx} \frac{- m_{\eta_{I}}^{2} + m_{\eta_{R}}^{2}}{v_1^2} {\,,} \\
			\lambda_9 &~{\approx} \frac{-2 \mu _{sb}^2 + m_{\eta_{I}}^{2} + m_{\eta_{R}}^{2} - v_\sigma^2 \lambda_{12} }{v_1^2} {\,,} \\
			\lambda_{13} &~{\approx} \frac{m_{\left( \varphi \right)_{R}}^2-m_{\left( \varphi \right)_{I}}^2}{2v_\sigma^2} {\,.}
		\end{aligned}
	\end{equation}	
	{Expressions for $\lambda_1$ and $\lambda_4$ are exact, whereas all others are valid in the approximation $\lambda_{14}, \lambda_{15} \rightarrow 0$.}	
	\subsection{Implications for collider physics and Dark Matter phenomenology}
	
	Finally, to close this section we wish to provide a brief discussion of the key phenomenological implications of the multi-scalar sector in the considered model. It is expected that the presence of electrically charged scalars in the inert doublet will provide an extra contribution to the Higgs diphoton decay rate $h\to \gamma\gamma$, which will yield a deviation of the Higgs diphoton signal strength from the SM expectation. However, that deviation is expected to occur in the experimentally allowed range in a large region of the model parameter space. Besides, given that singlet scalar field acquires a VEV at the TeV scale, its mixing with the CP-even part of the active doublet is suppressed implying that the couplings of the $125$ GeV SM like Higgs boson will be very close to the SM expectation realising the so-called Higgs alignment limit. Thus, such a scenario appears to be consistent with the SM in such a limit, at least, at the leading order (LO) level.
	
	It is generally expected that the triple Higgs coupling can be affected compared to its SM value in typical multi-scalar extensions of the SM. In our model, the trilinear Higgs coupling, $\lambda_{hhh}$, deviates from the SM prediction through a mixing with the $\sigma$ field. In particular, at LO a simple analytical expression can be derived 
	\begin{equation}\label{eq:trilinear_higgs}
		\begin{aligned}
			\lambda^{\mathrm{LO}}_{hhh} = 6\cos^3\alpha_h v \lambda_1 + 6\sin^3\alpha_h v_\sigma\lambda_4 + 3 \cos\alpha_h \sin^2\alpha_h v \lambda_8 + 3\cos^2\alpha_h \sin\alpha_h v_\sigma \lambda_8 \,,
		\end{aligned}
	\end{equation}
	where $\alpha_h$ is the CP-even mixing angle between $h_1$ (Higgs) and $h_2$. It is not hard to notice that in the limit of $\alpha_h\rightarrow 0$, the tree-level SM Higgs coupling $\lambda_{hhh}^{\mathrm{SM}} = 6v\lambda_1$ is successfully reproduced as we effectively work very close to the alignment limit. Hence, the potential corrections to $\lambda^{\mathrm{LO}}_{hhh}$ are very suppressed and we expect no significant deviations from the SM expectation. On the other hand, previous works \cite{Basler:2019iuu,Bahl:2022jnx,Osland:2008aw} have suggested that one-loop corrections to the $\lambda_{hhh}$ can actually be quite sizable in some BSM extensions, resulting, in some cases, in an increase of order $\mathcal{O}(50\%)$ for certain combinations coupling/mass parameters. In our case, next-to-leading order (NLO) contributions will be dominant and represent the bulk of the new physics contribution to $\lambda_{hhh}$. So, for a correct understanding of the impact on $\lambda_{hhh}$, one must take these contributions into account. In this regard, we have computed all relevant one-loop contributions to the triple Higgs coupling using the formalism of Ref.~\cite{Camargo-Molina:2016moz}. Note that the formulas derived here are valid in the limit of zero external momentum, and therefore, only the contributions coming from the heavier states are relevant. In particular, we have considered all possible contributions coming from all physical scalars of the model ($\eta^\pm$, $h_1$, $h_2$, $\eta_R$, $\varphi_R$, $\sigma_I$, $\eta_I$ and $\varphi_I$), as well as the loop diagram involving the virtual exchange of the top quark. The loops containing the gauge bosons and the lighter scalar bosons were neglected, as they are expected to be subdominant. The full one-loop formulas are listed in Appendix~\ref{app:one-loop}. In what follows, in our numerical analysis we will take into account the existing constraints on $\lambda_{hhh}$ available from the LHC \cite{ATLAS-CONF-2022-050,ATLAS-CONF-2021-052}. Future collider measurements \cite{ATL-PHYS-PUB-2022-018,Homiller:2018dgu} can further probe possible deviations of $\lambda_{hhh}$ from the SM value.

\begin{figure*}[tbh]
\begin{center}
\includegraphics[scale=0.29]{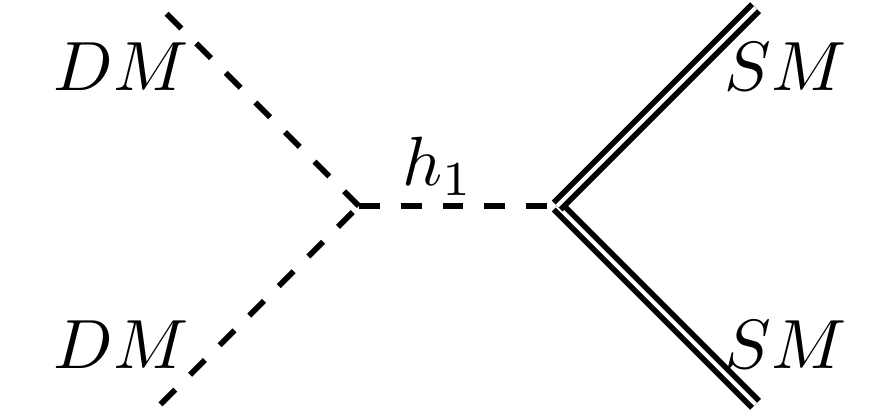}
  \includegraphics[scale=0.29]{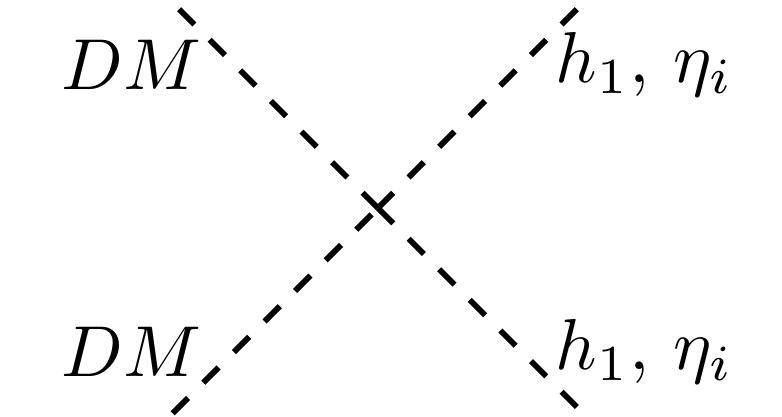}
  \includegraphics[scale=0.29]{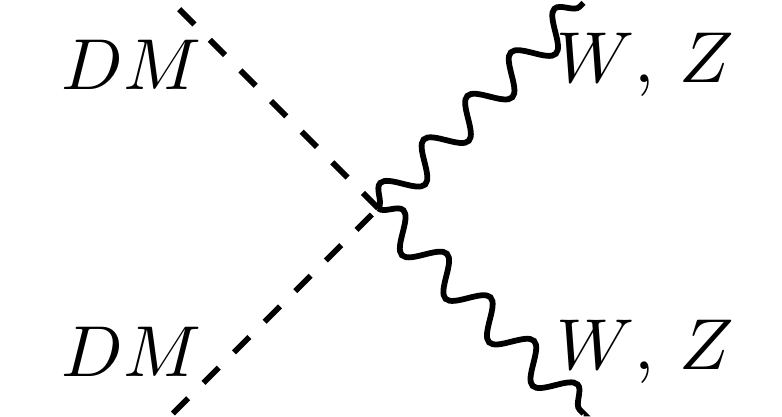}
  \includegraphics[scale=0.29]{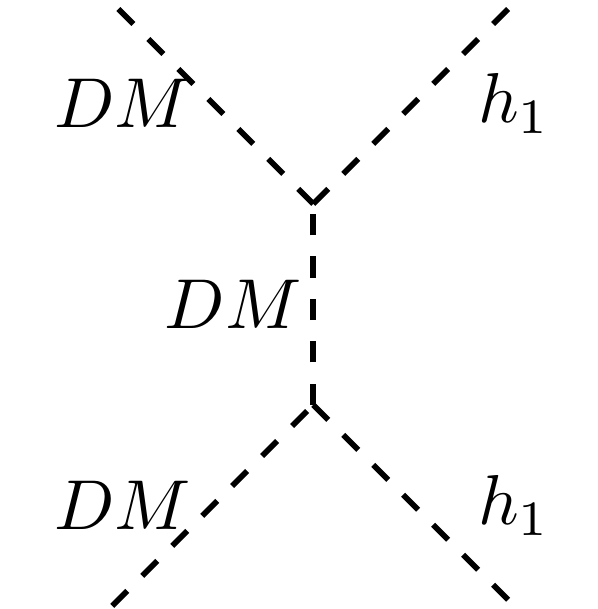}
\end{center}
\caption{Dominant Feynman diagrams contributing to the DM annihilation for the scenario considered.}
\label{figDMFeyn}
\end{figure*} 

\begin{figure}[th]
		\captionsetup{justification=raggedright}
		\begin{center}
			\includegraphics[width=0.7\textwidth]{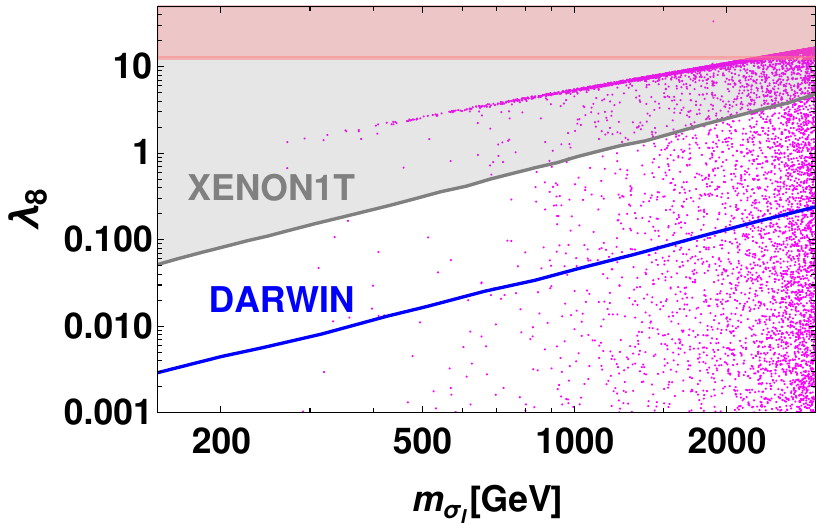}
		\end{center}
		\caption{Parameter space that gives the correct relic density in the $m_{\sigma_I}-\lambda_8$ plane. The pink band is disfavored by perturbativity.
  The gray region is excluded by XENON1T~\cite{XENON:2018voc} while the blue line corresponds to the sensitivity of DARWIN~\cite{DARWIN:2016hyl}.}
		\label{DMfig}
	\end{figure}

Another interesting implication of our model is for understanding the structure of DM. The Majoron $\sigma_I$, which is the pseudo-Nambu-Goldstone boson associated with the spontaneous breaking of the global lepton number symmetry, can be identified as the WIMP DM candidate. Its relic density can be generated by the standard freeze-out mechanism. Assuming for simplicity that the cosmological DM is dominated by such a scalar DM candidate, it would annihilate mainly into $WW$, $ZZ$, $t\overline{t}$, $b\overline{b}$ and $h_1h_1$ channels in the early Universe via a Higgs portal scalar interaction. These interactions will contribute to the DM relic density, which can be accommodated for appropriate values of the scalar DM mass, which in most cases are at the TeV scale. Such a scalar DM candidate would also scatter off a nuclear target in a detector via Higgs boson exchange in the $t$-channel, giving rise to a constraint on the Higgs portal scalar interaction coupling. In the low mass region, the invisible Higgs decay constraints should be taken into account and would imply a lower bound for the scalar DM candidate mass of about $60$ GeV. Besides that, in a small window around half of the SM Higgs boson mass, the DM relic density constraints will be successfully accounted for. In Fig.\ref{DMfig}, we have plotted the parameter space that gives the correct relic density in the $m_{\sigma_I}-\lambda_8$ plane, where $\lambda_8$ is the Higgs portal coupling. In calculating the relic density, we have assumed for simplicity that the dominant annihilation channels of the DM are the ones into $WW$, $ZZ$, $h_1 h_1$, $t\overline{t}$, $b\overline{b}$, and the components of the doublet 
$\eta$. The corresponding Feynman diagrams are given in Fig.~\ref{figDMFeyn}. The cross sections for the relevant annihilation channels are given as \cite{Bhattacharya:2016ysw},
\begin{eqnarray}
v_{rel}\sigma \left( \sigma_I \sigma_I \to WW\right)  &=&\frac{\lambda_8}{32\pi }\frac{s\left( 1+\frac{12m_{W}^{4}}{s^{2}}-\frac{4m_{W}^{2}}{s}\right) }{\left( s-m_{h_1}^{2}\right)^{2}+m_{h_1}^{2}\Gamma_{h_1}^{2}}\sqrt{1-\frac{4m_{W}^{2}}{s}}, \\
v_{rel}\sigma \left( \sigma_I \sigma_I \to ZZ\right)  &=&\frac{\lambda_8 }{64\pi }\frac{s\left( 1+\frac{12m_{Z}^{4}}{s^{2}}-\frac{4m_{Z}^{2}}{s}\right) }{\left( s-m_{h_1}^{2}\right)^{2}+m_{h_1}^{2}\Gamma_{h_1}^{2}}\sqrt{1-\frac{4m_{Z}^{2}}{s}}, \\
v_{rel}\sigma \left( \sigma_I \sigma_I \to q\overline{q}\right)  &=&
\frac{N_{c}\lambda_8^{2}m_{q}^{2}}{16\pi }\frac{\sqrt{\left( 1-\frac{4m_{q}^{2}}{s}\right) ^{3}}}{\left( s-m_{h_1}^{2}\right) ^{2}+m_{h_1}^{2}\Gamma_{h_1}^{2}}, \\
v_{rel}\sigma \left( \sigma_I \sigma_I \to h_1 h_1\right)  &=&\frac{\lambda_8^{2}}{64\pi s}\left( 1+\frac{3m_{h_1}^{2}}{s-m_{h_1}^{2}}-\frac{2\lambda_8
v^{2}}{s-2m_{h_1}^{2}}\right) ^{2}\sqrt{1-\frac{4m_{H_1}^{2}}{s}},\\
v_{rel}\sigma \left( \sigma_I \sigma_I \to \eta_i \eta_i\right)  &=&\frac{\lambda_{10} ^{2}}{64\pi s}\sqrt{1-\frac{4m_{\eta_i}^{2}}{s}},
\end{eqnarray}%
where $\sqrt{s}$ is the centre-of-mass energy, $N_{c}=3$ stands for the color factor, $m_{h_1}=125.7$ GeV and $\Gamma _{h_1}$ is the total decay width of the SM Higgs boson which is equal to 4.1 MeV. From the above scattering cross sections, the current DM relic abundance in the Universe can be obtained as (c.f. Ref.~\cite{Edsjo:1997bg,ParticleDataGroup:2022pth}),
\begin{equation}
\Omega h^{2} = \frac{0.1~~\textrm{pb}}{\langle \sigma v \rangle },\,%
\hspace{1cm}\langle \sigma v \rangle =\frac{A}{n_{eq}^{2}},
\end{equation}%
where $\langle \sigma v \rangle $ is the thermally averaged
annihilation cross section, $A$ is the total annihilation rate per unit
volume at temperature $T$ and $n_{eq}$ is the equilibrium value of the
particle density, which are given as~\cite{Edsjo:1997bg},
\begin{eqnarray}
A &=&\frac{T}{32\pi ^{4}}\int_{4m_{\sigma_I}^{2}}^{\infty}\sum_{p=W,Z,t,b,H_1,\eta_i} g_{p}^{2}\frac{s\sqrt{s-4m_{\sigma_I}^{2}}}{2}%
v_{rel}\sigma \left( \sigma_I \sigma_I \rightarrow pp\right)
K_{1}\left( \frac{\sqrt{s}}{T}\right) ds,  \notag \\
n_{eq} &=&\frac{T}{2\pi ^{2}}\sum_{p=W,Z,t,b,H_1,\eta_i}g_{p}m_{\sigma_I}^{2}K_{2}\left( \frac{m_{\sigma_I}}{T}\right) ,
\end{eqnarray}
with $K_{1}$ and $K_{2}$ being the modified Bessel functions of the second kind of order 1 and 2, respectively.  
We have taken $T=m_{m_{\sigma_I}}/20$ following Ref.~\cite{Edsjo:1997bg}.  
The DM relic density thus determined should match the required value ~\cite{Planck:2018vyg},
\begin{equation}
\Omega _{DM}h^{2}= 0.1200\pm 0.0012.
\label{Omegavalue}
\end{equation}

In generating Fig.~\ref{DMfig}, we have varied $\lambda_8$ from 0 to $4\pi$ and the masses of $\eta_R$, $\eta_I$ and $\eta^{\pm}$ in the range 0.2 - 5 TeV. In this figure, the pink band is disfavored by perturbativity bounds.
  The gray region is excluded by the constraints from XENON1T~\cite{XENON:2018voc} whereas the blue line corresponds to the sensitivity of DARWIN~\cite{DARWIN:2016hyl}. From this figure, one can see that most of the allowed parameter space corresponds to a dark matter mass in the TeV range though there are a few allowed points in the sub-TeV region as well.

	\section{Neutrino mass and lepton flavor violation}\label{sec:observables}
	
	Once $\Phi$ and $\sigma$ develop VEVs, a non-trivial contribution to the neutrino mass matrix is generated at two-loop level, whose diagram is illustrated in Fig.~\ref{fig:Neutrinoloopdiagram}. The active light neutrino mass matrix in this case is given as,
	\begin{equation}\label{eq:neutrino_mass}
		\begin{aligned}
			\left( M_{\nu }\right) _{ij} = &\frac{
				\left( y_{N}\right) _{in}\left( y_{\Omega }^{\ast }\right) _{nk}\left(
				y_{\Omega }^{\dag }\right) _{kr}\left( y_{N}^{T}\right) _{rj}m_{\Psi _{k}}}{
				4(4\pi )^{4}}\\ &\int_{0}^{1}d\alpha \int_{0}^{1-\alpha }d\beta \frac{1}{\alpha
				(1-\alpha )}\Biggl[G\left( m_{\Psi _{k}}^{2},m_{RR}^{2},m_{RI}^{2}\right)
			-G\left( m_{\Psi _{k}}^{2},m_{IR}^{2},m_{II}^{2}\right) \Biggr] \,, 
		\end{aligned}
	\end{equation}
	where $n,k,r = 1,2$. Here, the loop integral $I$ can be written as \cite{Kajiyama:2013rla} 
	\begin{eqnarray}
		G(x^{2},y^{2},z^{2})\!\!\! &=&\!\!\!\frac{x^{2}y^{2}\log
			\left( \displaystyle\frac{y^{2}}{x^{2}}\right)
			+y^{2}z^{2}\log \left( \displaystyle\frac{z^{2}}{y^{2}}
			\right) +z^{2}x^{2}\log \left( \displaystyle\frac{x^{2}}{
				z^{2}}\right) }{(x^{2}-y^{2})(x^{2}-z^{2})} \,,  \notag \\
		m_{ab}^{2}\!\!\! &=&\!\!\!\frac{\beta m_{\left( \eta \right)
				_{a}}^{2}+\alpha m_{\left( \varphi \right) _{b}}^{2}}{\alpha (1-\alpha )}
		\quad (a,b=R:\mathrm{or}:I) \,.
	\end{eqnarray}
	
	Note that the lightest active neutrino is massless which implies that there is only one Majorana phase in the PMNS matrix. A convenient way to incorporate the bounds from the neutrino oscillation data is to express the Yukawa coupling matrix using the Casas-Ibarra parametrization. For the model considered in this paper, one can express $y_{N}$ as, 
	\begin{equation}\label{eq:yN_cI}
		y_N = U_{\mathrm{PMNS}}^* \sqrt{M_\nu^{\mathrm{diag}}} R^T A^{-1}\,,
	\end{equation}
	where $U_{\rm PMNS}^*$ is the $3\times 3$ PMNS neutrino mixing matrix, $M_\nu^{\mathrm{diag}} = ~\text{diag} (m_1, m_2, m_3)$ is the diagonal neutrino mass matrix, $R$ is a $2\times 3$ matrix that is given as, 
	\begin{equation*}
		R= 
		\begin{pmatrix}
			0 & \text{cos} z & -\text{sin} z \\ 
			0 & \text{sin} z & \text{cos} z%
		\end{pmatrix}
		~~~~~ \text{for normal hierarchy (NH) of light neutrino masses}~~ (m_1=0)~~~%
		\text{and}
	\end{equation*}
	\begin{equation*}
		R= 
		\begin{pmatrix}
			\text{cos} z & -\text{sin} z & 0 \\ 
			\text{sin} z & \text{cos} z & 0%
		\end{pmatrix}
		~~~~~ \text{for inverted hierarchy (IH) of light neutrino masses}~~ (m_3=0).
	\end{equation*}
	The matrix $A$ is given by
	\begin{equation*}
		A = y_\Omega^* \Lambda y_\Omega^\dag \,,
	\end{equation*}
	where $\Lambda = 
	\begin{pmatrix}
		\Lambda_1 & 0 \\ 
		0 & \Lambda_2%
	\end{pmatrix}
	$ with, 
	\begin{equation*}
		\Lambda_k = \frac{m_{\Psi _{k}}}{4(4\pi )^{4}} \int_{0}^{1}d\alpha
		\int_{0}^{1-\alpha }d\beta \frac{1}{\alpha (1-\alpha )}\Biggl[G\left(
		m_{\Psi _{k}}^{2},m_{RR}^{2},m_{RI}^{2}\right) -G\left( m_{\Psi
			_{k}}^{2},m_{IR}^{2},m_{II}^{2}\right) \Biggr] \,.
	\end{equation*}
	Under this parameterization, the experimentally observed neutrino mass differences and the PMNS mixing matrix can be given as input in the numerical scan, with the Yukawa coupling being found as output. Since we do not have control over the magnitude of the Yukawa couplings, we only allow solutions which respect perturbativity, i.e. $\abs{Y_N}, \abs{Y_\Omega} < \sqrt{4\pi}$. For the purpose of numerical scans, we only consider the normal ordering scenario for the neutrino masses (that is, $m_1 = 0$).
	
	The presence of the Majorana fermions can induce LFV decays, such as $\mu \rightarrow e \gamma$, which are strongly constrained by experiment. In our model, such decays are mediated at one-loop level via virtual exchanges of the neutral fermions and the charged scalars. The branching fraction for the two-body decay process $\ell_i \rightarrow \ell_j \gamma$, where $i = e,\mu,\tau$ is given as \cite{Ma:2001mr, Toma:2013zsa, Vicente:2014wga, Lindner:2016bgg}
	\begin{equation}\label{eq:mu_to_egamma}
		\mathrm{BR}\left( l_{i}\rightarrow l_{j}\gamma \right) =\frac{3\left( 4\pi \right)
			^{3}\alpha _{\mathrm{em}}}{4G_{F}^{2}}\left\vert \frac{x_{is}^{\left(
				\nu \right) }x_{js}^{\left( \nu \right) }}{2\left( 4\pi \right) ^{2}m_{\eta
				^{\pm }}^{2}}F\left( \frac{m_{N_{sR}}^{2}}{m_{\eta ^{\pm }}^{2}}\right)
		\right\vert ^{2}\mathrm{BR}\left( l_{i}\rightarrow l_{j}\nu _{i}\overline{\nu _{j}}%
		\right) \,,  
	\end{equation}
	with $s=1,2$. Here, $\alpha_{\mathrm{em}} = 1/137$ is the fine-structure constant, $x_{is}^{\left( \nu \right) }=\sum_{k=1}^{3}(y_N)_{ks}^{\left( \nu \right)}\left( V_{lL}^{\dagger }\right)_{ik}$, with $V$ being the left-handed charged lepton mixing matrix, $G_F = 1.166364\times 10^{-5}~\mathrm{GeV^{-2}}$ is the Fermi constant, which in our case is the identity matrix, $m_{\varphi ^{\pm }}$ are the masses of the charged scalar components of the $\mathrm{SU(2)_{L}}$ inert doublet $\varphi$, and $m_{N_{sR}}$ ($s=1,2$) correspond to the masses of the right-handed Majorana neutrinos $N_{sR}$, which due to the charge assignment, are generated at one-loop level. The loop function $F$ is written as,
	\begin{equation}\label{eq:loop_function}
		F\left( x\right) = \frac{1-6x+3x^{2}+2x^{3}-6x^{2}\ln x}{6\left( 1-x\right)} \,.
	\end{equation}
    {The masses of the right handed neutrinos $N_{sR}$ are given as,
    \begin{equation}
    m_{N} = \frac{{y_\Omega}^2 m_\psi}{8 \pi^2}\Bigg[\frac{m_{\varphi_R}^2}{m_{\varphi_R}^2-m_\psi^2}\ln(\frac{m_{\varphi_R}^2}{m_{\psi}^2}) -    \frac{m_{\varphi_I}^2}{m_{\varphi_I}^2-m_\psi^2}\ln(\frac{m_{\varphi_I}^2}{m_{\psi}^2})\Bigg],
    \end{equation}
    where we have taken $y_\Omega$ and $m_\psi$ to be diagonal and degenerate for simplicity.}

	The most stringent bounds for LFV come from muon decay measurements, namely, $\mu\rightarrow e \gamma$. Current experimental results put an upper bound on the branching ratio, which reads as $\mathrm{BR}\left(\mu \rightarrow e\gamma \right) < 4.2 \times 10^{-13}$ \cite{MEG:2016leq}. The model can also give rise to $\mu \rightarrow e$ conversion in the atomic nucleus. But the existing bounds are not as restrictive as the ones from $\mathrm{BR}\left(\mu \rightarrow e\gamma \right)$. However, among various planned experiments, the $\mu \rightarrow e$ conversion in the Al nuclei has the best projected sensitivity of $\sim 10^{-17} $ \cite{Bernstein:2013hba}. The branching ratio for this process can approximately be expressed as \cite{Lindner:2016bgg},
	\begin{equation}
		\mathrm{CR} (\mu \, \mathrm{Al} \rightarrow e \, \mathrm{Al}) \approx \frac{1}{350} \mathrm{BR}\left(\mu \rightarrow e\gamma \right) \,.
	\end{equation}
	\begin{figure}[th]
		\captionsetup{justification=raggedright}
		\begin{center}
			\includegraphics[width=0.7\textwidth]{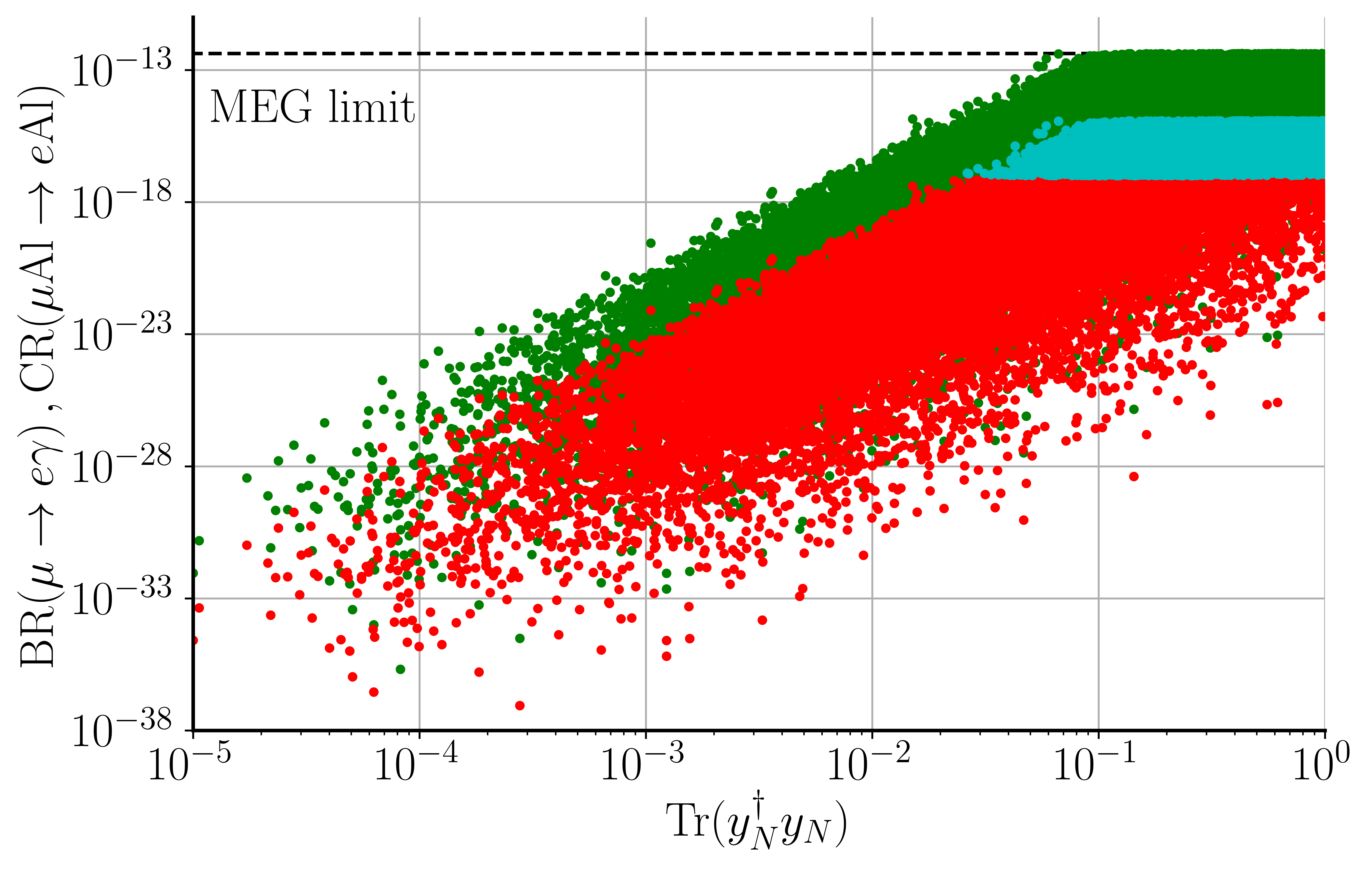}
		\end{center}
		\caption{The predictions for $\mathrm{BR}\left(\mu \rightarrow e\gamma \right)$ (green points) and $\mathrm{CR} (\mu \, \mathrm{Al} \rightarrow e \, \mathrm{Al})$ (red points) as functions of Tr$[y_N^\dag y_N]$. All the points shown are within the existing experimental bounds marked by the horizontal dashed line. The cyan points correspond to the region within the projected sensitivity for $\mu \, \mathrm{Al} \rightarrow e \, \mathrm{Al}$ conversion.}
		\label{fig:muegamma}
	\end{figure}
	
	In this regards, in Fig.~\ref{fig:muegamma} shows the predictions for $\mathrm{BR}\left(\mu \rightarrow e\gamma \right)$ (green points) and $\mathrm{CR} (\mu \, Al \rightarrow e \, Al)$ (red points) as functions of Tr$[y_N^\dag y_N]$. Only the values that are within the existing bounds are shown, whose limit is indicated by a dashed horizontal line in the plot. The cyan points correspond to the region within the projected sensitivity for $\mu \, \mathrm{Al} \rightarrow e \, \mathrm{Al}$ conversion.
	
	For all the points shown in Fig.~\ref{fig:muegamma}, neutrino oscillation data is respected, with the relevant experimental parameters (namely, neutrino mass differences and the PMNS mixing angles) being given as input and allowed to vary within their respective $3\sigma$ uncertainties \cite{Capozzi:2017ipn,Esteban:2020cvm,deSalas:2020pgw}. The Majorana phase is also varied in the range $[0,\pi]$ and $y_\Omega$ is taken to be diagonal where the entries are allowed to vary in the range $[0,1]$. Here, $z$ is defined as $x_1 + y_1i$ with $x_1$ and $y_1$ being varied in the range $[-2, +2]$, the fermions $\Psi_{kR}$ are taken to be degenerate, with their masses varied in the range $[0.5,5]$ TeV. For the scalar potential parameters, we employ the inverted equations in \eqref{eqn:inverted_eqs} to allow for the physical masses to be given as input. In this regard, we have considered the following parameter ranges:
	\begin{itemize}
		\item $m_{\eta_{I}}, m_{\varphi_{I}} \rightarrow [0.5,3.2]$ TeV, $m_{\eta_{R}} \rightarrow [m_{\eta_{I}}, m_{\eta_{I}}+5 \,\textrm{GeV}]$, $m_{\varphi_{R}} \rightarrow [m_{\varphi_{I}}, m_{\varphi_{I}}+5 \,\textrm{GeV}]$, $m_{\eta ^{\pm }} \rightarrow [m_{\eta_{I}}-5 \,\textrm{GeV}, m_{\eta_{I}}]$
		\item  $v_1 = 246$ GeV, $m_{h_2} \rightarrow [0.5,5]$ TeV, $v_\sigma \rightarrow [5,30]$ TeV, 
		\item $\lambda_{2,5,8,9,10,11,12} \rightarrow [0,1]$. The remaining $\lambda$'s are treated as outputs in the inversion procedure. Notice that we have inverted the expressions for $\lambda_5$ and $\lambda_9$, which can be solved such that $\mu_{\eta }^{2}$ and $\mu_{\varphi }^{2}$ are instead given as output parameters, where we have taken them to be both greater than $0$. Also, in the parameter space that has been considered, we have ensured that all the quartic couplings, $\lambda_i > 0$, even though this is only a sufficient condition but not necessary.
	\end{itemize}
	
	\section{Phase transitions and primordial gravitational waves}\label{sec:GWs_pt}
	
	One of the more interesting phenomenological consequences of extended scalar sectors compared to the SM framework is the possibility for the existence of FOPTs. Indeed, such a phenomenon is not present in the SM (both the EW and QCD phase transitions are of second order), making it pure beyond-SM driven (BSM) physics. Such transitions can lead to the generation of PGWs. With the detection of GWs at LIGO \cite{Abbott:2016blz}, a new era of multi-messenger astronomy has arisen, where such events may give us unique perspectives into new physics beyond the SM. It has been recently proposed that the cosmological phase transitions between different vacua at finite temperatures (such as those associated with symmetry breaking in BSM models) may give rise to gravitational imprints that should be observable in future ground- and space-based experiments \cite{Greljo:2019xan,King:2020hyd}. The PGWs signals can also provide hints into the nature of the neutrino masses in the context of distinct seesaw mechanisms \cite{Addazi:2019dqt}.
	
	As was mentioned above, PGWs can arise from FOPTs, which produce a stochastic GW background in the early Universe which is formed out of equilibrium due to fast-expanding vacuum bubbles \cite{Kosowsky:1992vn}. These bubbles eventually collide and merge to give rise to the GW echoes \cite{Hashino:2018wee}. The characteristics of the phase transition such as its strength and inverse duration are determined by the structure of the effective potential at finite temperatures and are highly sensitive, in particular, to the properties of the potential barrier between the phases and to the critical temperature of the transition (see e.g.~Refs.~\cite{Morais:2019fnm,Greljo:2019xan,Cline:2021iff,Freitas:2021yng,Zhang:2021alu,DiBari:2021dri} and references therein). The phenomenon of strong FOPTs is quite generic for multi-scalar extensions of the SM (in particular, for those that undergo several symmetry breaking steps) such as the one we address in this article. Therefore, PGWs emerging from cosmological FOPTs are often considered as a potentially useful source of phenomenological information about multi-scalar extensions of the SM which may be complementary to constraints coming from collider searches.
	
	In order to fully classify and study the PGW spectra emergent in our model, there are a total of five main ingredients. The first one is inherently dependent on the considered model structure -- the finite temperature effective potential \cite{Dolan:1973qd,Weinberg:1974hy,Brandenberger:1984cz}. Generically, it can be written as \cite{Quiros:1999jp,Curtin:2016urg}
	\begin{equation}\label{eq:effective_potential}
		V_{\mathrm{eff}} = V^{(0)} + V_{\mathrm{CW}}^{(1)} + V_{T\neq 0} + V_{\mathrm{CT}} \,,
	\end{equation}
	where the first term, $V^{(0)}$, is the tree-level scalar potential given in Eq.~\eqref{eq:scalar_potential}. The second term is the Coleman-Weinberg one-loop correction,
	\begin{equation}\label{eq:one-loop-level}
		V^{(1)}_{\mathrm{CW}} = \sum_{a} (-1)^F n_a \frac{m_a^2(\phi_b)}{64\pi^2} \Bigg[\ln(\frac{m_a^2(\phi_b)}{\mu^2}) - C_a\Bigg] \,,
	\end{equation}
	where $F$ is 0 for bosons and 1 for fermions, $m_a(\phi_b)$ is the $\phi_a$-field dependent mass of the particle $a$, $n_a$ is the number of degrees of freedom for each particle $a$ and $\mu$ is a renormalisation scale. The particle degrees of freedom can be computed as $(-1)^{2s}QN(2s+1)$, where $s$ is the spin of the particle, $Q = 1(2)$ for neutral (charged) particles, and $N$ is the number of colours. In this work, we consider the $\overline{\mathrm{MS}}$ renormalisation scheme such that we have $C_a=3/2$ for scalars, fermions and longitudinally polarized gauge bosons, and $C_a=1/2$ for transverse gauge bosons. Additionally, the renormalisation scale is fixed as $\mu = \prod_{i=1}^{n} m_{\phi_i}^{1/n}$ where $i$ runs over all BSM scalars in the model. Since we allow to vary the masses of the scalar fields, $\mu$ will be different for each sampled point.
	
	The third term encodes thermal corrections to the potential,
	\begin{equation}\label{eq:thermal_corrections}
		V_{T\neq 0} = \frac{T^4}{2\pi^2}\sum_{a} (-1)^Fn_a J_a\qty[\frac{m_a^2(\phi_a)}{T^2}] \,,
	\end{equation}
	where $T$ is the temperature, $J_0$ (for bosons) and $J_1$ (for fermions) are the standard thermal integrals found e.g.~in Ref.~\cite{Quiros:1999jp}. At the leading $(m/T)^2$ order in thermal expansions of $J_a$, Eq.~\eqref{eq:thermal_corrections} can be approximated as 
	\begin{equation}\label{eq:approx_DVT}
		\Delta V(T) \simeq \frac{T^4}{2\pi^2} \qty(\Tr{M^2(\phi_\alpha)} + \sum_{i=W,Z,\gamma} n_i m_i^2(\phi_\alpha) + \sum_{i=f_i} \frac{n_i}{2}m_i^2(\phi_\alpha)) \,,
	\end{equation}
	where in the last sum, all the fermions of the model are included. Here, $M$ is the field-dependent Hessian matrix in the classical field approximation where all fields are fixed to their classical (background) components, that is
	\begin{equation}\label{eq:cfield_aprox}
		\Phi \rightarrow 
		\begin{pmatrix}
			0 \\ v_{\Phi}/\sqrt{2}	
		\end{pmatrix}\,, \quad
		\sigma \rightarrow 
		v_{\sigma}\,, \quad
		\eta \rightarrow 
		\begin{pmatrix}
			0 \\ 0	
		\end{pmatrix}\,, \quad
		\varphi \rightarrow 0 \,,
	\end{equation}
	and the degrees of freedom, $n_i$, for different states are as follows \cite{Quiros:1999jp}: for quarks $n_a = 12$, for charged leptons $n_a = 4$, for neutral leptons $n_a = 2$, for the gauge bosons $n_W = 6$, $n_Z = 3$, $n_\gamma = 2$ and for the longitudinal component of the photon $n_{\gamma_L} = 1$. Additional higher-order corrections must also be included in the form of Daisy (ring) diagrams \cite{Dolan:1973qd,Parwani:1991gq,Arnold:1992rz,Espinosa:1995se}. These can be parameterized as thermal corrections to the quadratic mass terms of the bare potential and read as
	\begin{equation}\label{eq:thermal}
		\mu^2(T) \rightarrow \mu^2 + c_{\phi}(T^2) \equiv \mu^2 + \frac{\delta^2 \Delta V(T, \phi_a)}{\delta \phi_a} \,.
	\end{equation} 
	For the considered model, we have
	\begin{equation}\label{eq:corrections_mus}
		\begin{aligned}
			&c_\phi(T^2) = \frac{T^2}{24} \Big[3g^2 + \frac{3}{2}(g^2 + g_Y^2) + 12(y_b^2 + y_t^2) + 4y_\tau^2 + 12\lambda_1 + 4\lambda_5 + 2(\lambda_6 + \lambda_8 + \lambda_9)\Big] \\
			&c_\sigma(T^2) = \frac{T^2}{24}(4(\lambda_{10} + \lambda_8) + 2\lambda_{12} + 8\lambda_4) \,,
		\end{aligned}
	\end{equation}
	where $g$ ($g_Y$) are the SU(2) (U(1)) gauge couplings and $y_i$ are the Yukawa couplings for each of the SM-like third-generation fermions. The gauge boson masses also gain thermal corrections and, since there are no new gauge bosons in the model, the formula is identical to what is found in the previous literature (see e.g.~Ref.~\cite{Quiros:1999jp}).
	
	The last relevant part of the effective potential in Eq.~\eqref{eq:effective_potential} is the counter-term potential $V_{\mathrm{CT}}$. The counter-terms are defined such that the minima of the tree-level potential remains fixed at zero temperature (i.e.~the tree-level masses remain the same at one-loop level), which is guaranteed by imposing that the first and second derivatives of the counter-term potential matches the first and second derivatives of the Coleman-Weinberg one-loop potential (for a pedagogical discussion, see for instance Ref.~\cite{Freitas:2021yng} and references therein). In our case, this results in the following conditions
	\begin{equation}\label{eq:counter_terms}
		\begin{aligned}
			& \delta \mu_\phi = \frac{3}{2v_\Phi} \frac{\partial V^{(1)}_{\mathrm{CW}}}{\partial h_1} - \frac{1}{2}\frac{\partial^2 V^{(1)}_{\mathrm{CW}}}{\partial h^2_1} + \frac{v_\sigma}{2v_\Phi}\frac{\partial^2 V^{(1)}_{\mathrm{CW}}}{\partial h_2 \partial h_1}\, , \quad\quad
			&&\delta \mu_{sb} = \frac{3}{4v_\sigma} \frac{\partial V^{(1)}_{\mathrm{CW}}}{\partial h_2} + \frac{v_\Phi}{4v_\sigma} \frac{\partial^2 V^{(1)}_{\mathrm{CW}}}{\partial h_2 \partial h_1} - \frac{1}{4}\frac{\partial^2 V^{(1)}_{\mathrm{CW}}}{\partial h_2^2}\, , \\
			& \delta \lambda_1 = \frac{1}{2v_\Phi^3} \qty(v_\Phi \frac{\partial^2 V^{(1)}_{\mathrm{CW}}}{\partial h_1^2}  - \frac{\partial V^{(1)}_{\mathrm{CW}}}{\partial h_1})\, ,&&\delta\lambda_4 = \frac{1}{2v_\sigma^3}\qty(v_\sigma \frac{\partial^2 V^{(1)}_{\mathrm{CW}}}{\partial h_2^2} - \frac{\partial V^{(1)}_{\mathrm{CW}}}{\partial h_2})\, , \\
			& \delta \lambda_8 = \frac{1}{v_\Phi v_\sigma} \frac{\partial V^{(1)}_{\mathrm{CW}}}{\partial h_2 \partial h_1}\,,
		\end{aligned}
	\end{equation}
	while for other parameters, we set $\delta\lambda_{i} = \delta\mu_{i} = 0$.
	
	The remaining four ingredients determine the FOPT dynamics, which include \cite{Hashino:2018wee}
	\begin{itemize}
		\item The inverse time-scale of the phase transition, $\beta$. This parameter essentially determines the bubble nucleation rate. Normalising it conventionally to the Hubble expansion rate, $H$, one writes 
		\begin{equation}\label{eq:beta_inv_time}
			\frac{\beta}{H} = T_n\frac{\partial}{\partial T}\qty(\frac{S(V_{\mathrm{eff}})}{T})\Bigr|_{\substack{T=T_n}} \,,
		\end{equation}
		where $S(V_\mathrm{eff})$ is the Euclidean action which depends on the structure of the effective potential in Eq.~\eqref{eq:effective_potential}.
		\item Nucleation temperature, $T_n$. It represents the temperature of the Universe directly after the phase transition, and can be computed by requiring that the probability for a single bubble nucleation per horizon volume is equal to one \cite{Moreno:1998bq}.
		\item Phase transition strength $\alpha$, which is related to the released latent heat during the phase transition. It is found in terms of the difference between the effective potential values before and after the transition via the following relation \cite{Hindmarsh:2017gnf,Hindmarsh:2015qta}
		\begin{equation}\label{eq:strength_anomaly}
			\alpha = \frac{30}{g_*\pi^2 T_n^2}\qty[V_{\mathrm{eff},i} - V_{\mathrm{eff},f} - \frac{T}{4}\qty(\frac{\partial V_{\mathrm{eff},i}}{\partial T} - \frac{\partial V_{\mathrm{eff},f}}{\partial T})] \,,
		\end{equation}
		being also normalised to the radiation energy density of the Universe at the nucleation temperature. Here, $g_*$ is the effective number of relativistic degrees of freedom in the cosmic plasma. The effective potential is evaluated at initial (metastable) phase, $V_{\mathrm{eff},i}$, and at the final (stable) phase, $V_{\mathrm{eff},f}$.
		\item Bubble wall velocity, $v_b$. This quantity represents the speed of the new phase at the interface of the nucleating bubble. In the current we will simply assume the case of supersonic detonations with $v_b = 0.95$. An improved analysis using the recently developed methods for the determination of the speed of sound \cite{Tenkanen:2022tly} and wall velocity \cite{Ai:2023see} is left for future work.
	\end{itemize}
	
	From this set of FOPT characteristics, the power spectrum associated with the production of PGWs can arise from three distinct sources,
	\begin{equation}\label{eq:power_spectrum}
		h^2\Omega_{\mathrm{GW}} = h^2\Omega_{\mathrm{BC}} + h^2\Omega_{\mathrm{SW}} + h^2\Omega_{\mathrm{TURB}} \,.
	\end{equation}
	Here, $h^2\Omega_{BC}$ is the power for GW spectrum produced via collision of vacuum bubbles and is a particularly important effect when such bubbles ``run-away'', that is, the friction created by the thermal plasma at the boundaries is not enough to stop the acceleration of the wall \cite{Hashino:2018wee}. A study has shown that this effect is subleading compared to the rest, and can usually be neglected \cite{Bodeker:2017cim}. The second term, $h^2\Omega_{\mathrm{SW}}$ is the power of GWs emerging from sound-waves triggered in the plasma once the bubble wall sweeps through the plasma at relativistic speeds. In this case, a vast portion of the released energy is converted to the waves' propagation in the plasma that surrounds the walls as sound waves \cite{Hashino:2018wee}. The last term is due to magnetohydrodynamic turbulence effects that arise due to non-linear effects on the sound waves in the plasma that are largely uncertain but expected to be less relevant for the considered FOPTs than the sound waves. For analytical expressions that connect each term of Eq.~\eqref{eq:power_spectrum} with the various parameters of the FOPTs introduced above, see \cite{Hashino:2018wee,LISACosmologyWorkingGroup:2022jok,Caprini:2019egz} and references therein.
	
	\subsection{Numerical results}
	
	We have implemented the thermal effective potential of the considered scotogenic two-loop neutrino mass model into the \texttt{CosmoTransitions} package \cite{Wainwright:2011kj} which enable us to compute numerically the bounce action and hence the probability for bubble nucleation at any given temperature. The key parameters that control the dynamics of the phase transitions, $\beta/H$ and $\alpha$, depend on derivative of the Euclidean effective action, hence, they can exhibit numerical instabilities if the action is not well-behaved. To minimize the corresponding numerical uncertainties, we perform an interpolation of the action on a point-by-point basis. The technique for the smoothing of the Euclidean action is described in detail in some of the author's previous work \cite{Freitas:2021yng}. Furthermore, for the scan, we have extended the ranges of various parameters to be more inclusive. Namely, we now perform a logarithmic scan over the quartic couplings in the ranges of $[10^{-8},\, 4\pi]$ and the masses of the BSM scalars to be above $200~\mathrm{GeV}$. Constraints from neutrino physics and LFVs are also imposed in the data points.
	
\begin{figure*}[htb!]
		\centering
		\begin{subfigure}[t]{0.29\textwidth}
			\centering
			\includegraphics[width=1.13\textwidth]{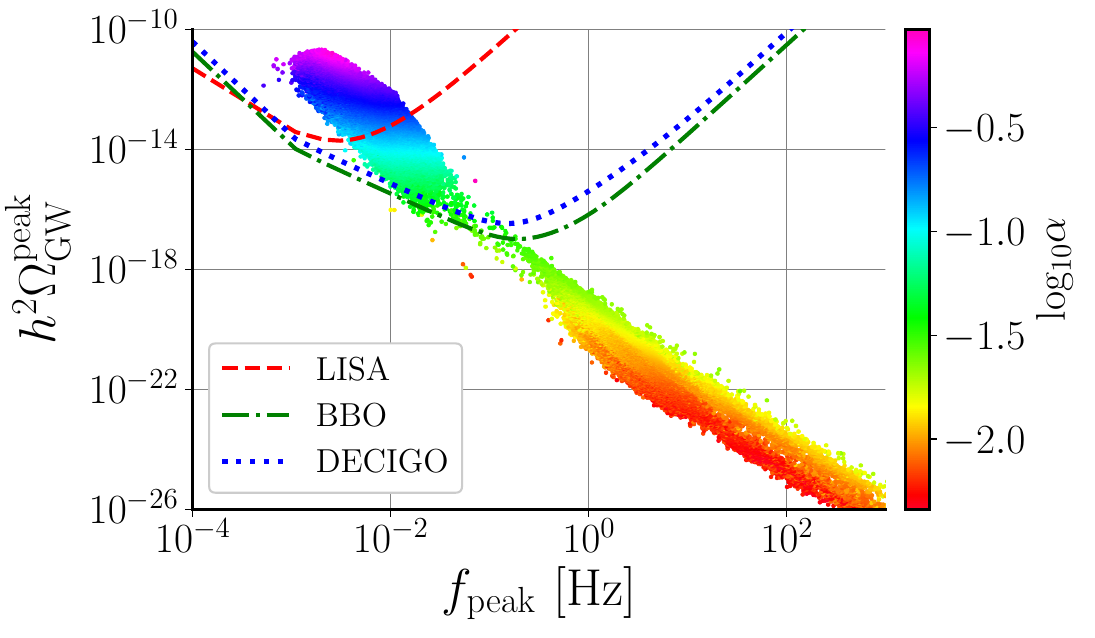}
			\caption{}
		\end{subfigure}
		\hfill
		\begin{subfigure}[t]{0.29\textwidth}
			\centering
			\includegraphics[width=1.01\textwidth]{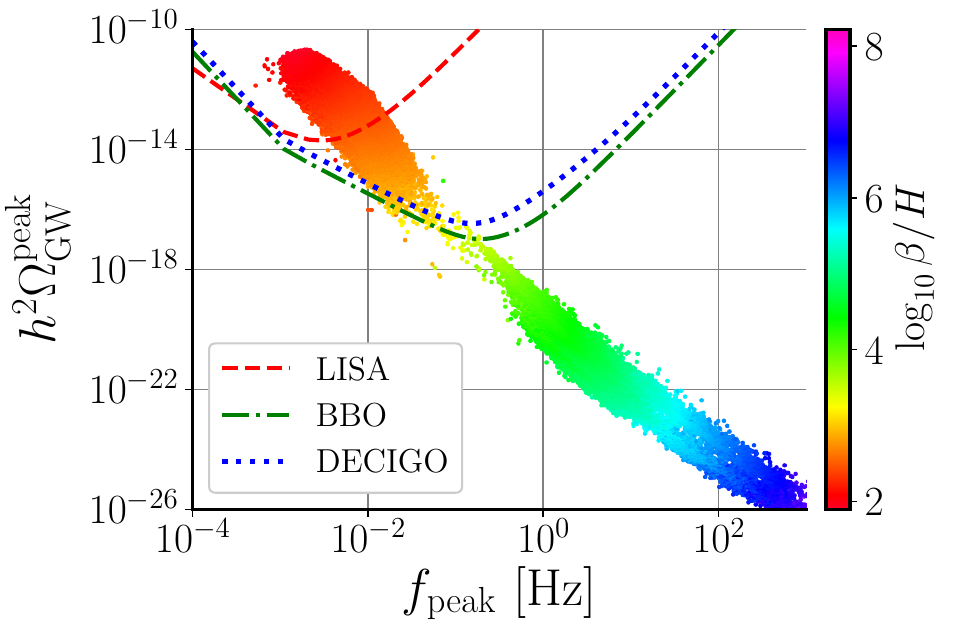}
			\caption{}
		\end{subfigure} 
		\hfill   
		\begin{subfigure}[t]{0.29\textwidth}
			\centering
			\includegraphics[width=1.07\textwidth]{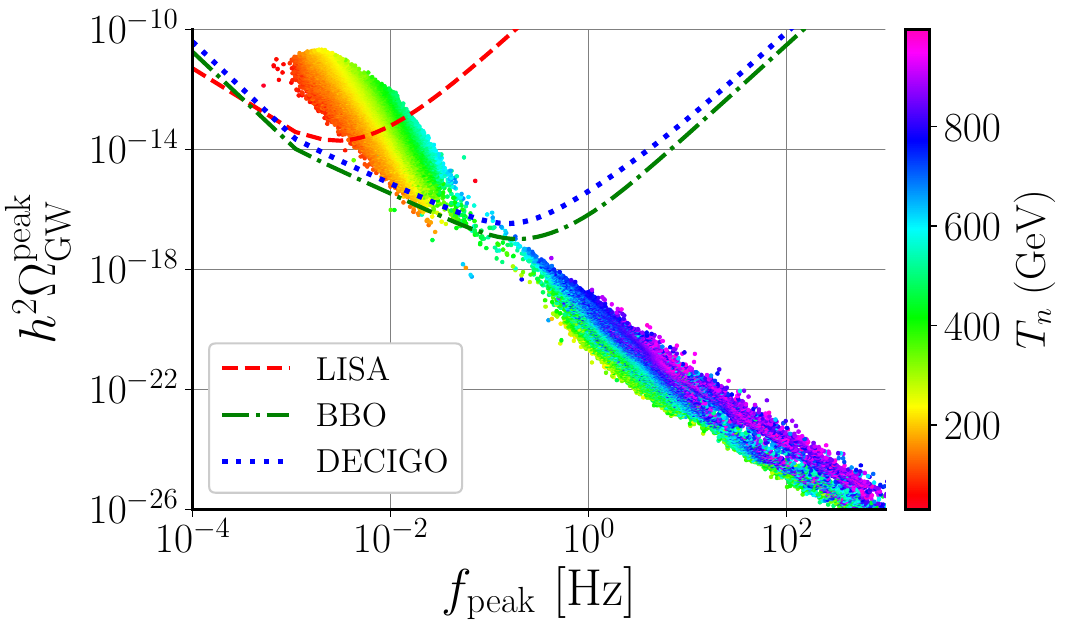}
			\caption{}
		\end{subfigure}
		\begin{subfigure}[t]{0.29\textwidth}
			\centering
			\includegraphics[width=1.10\textwidth]{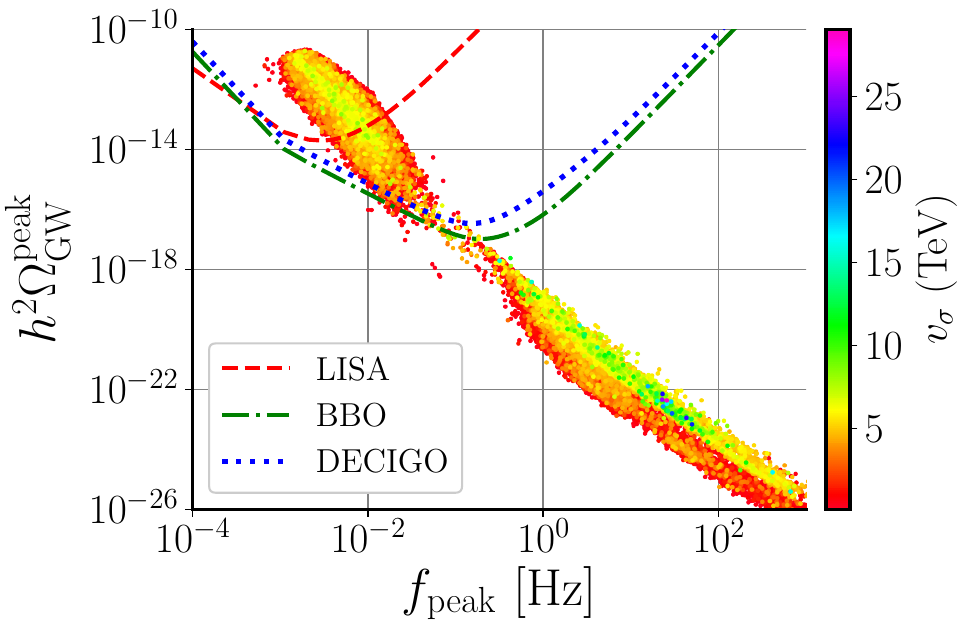}
			\caption{}
		\end{subfigure}
		\hfill
		\begin{subfigure}[t]{0.29\textwidth}
			\centering
			\includegraphics[width=1.10\textwidth]{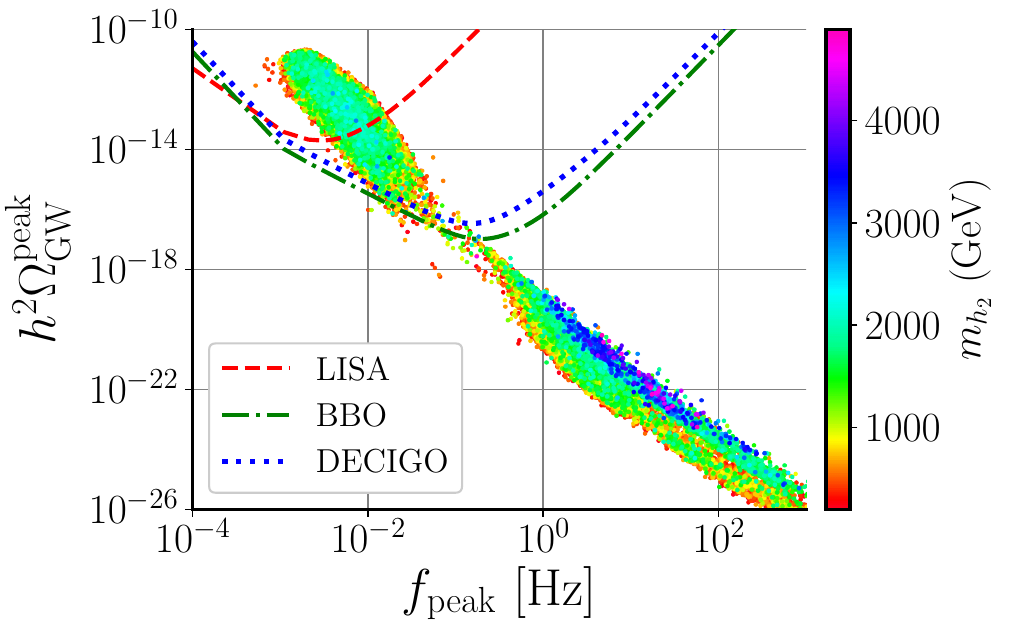}
			\caption{}
		\end{subfigure} 
		\hfill   
		\begin{subfigure}[t]{0.29\textwidth}
			\centering
			\includegraphics[width=1.10\textwidth]{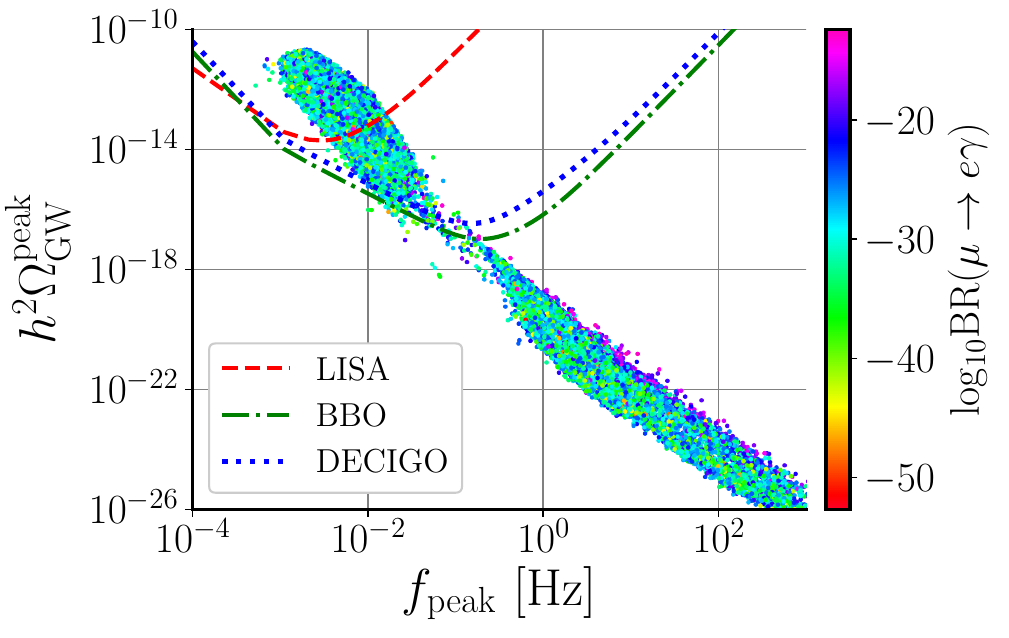}
			\caption{}
		\end{subfigure}
		\caption{Peak amplitude of the GW signal, $h^2 \Omega_{\mathrm{GW}}$ as a function of the peak frequency, $f_{\mathrm{peak}}$, in Hz. In colour, we have the logarithm of the phase transition strength $\alpha$ (a), the logarithm of the inverse time-scale of the phase transitions normalized to the Hubble parameter (b), the nucleation temperature in GeV (c), the VEV of the singlet $\sigma$ field in TeV (d), the mass of the second heaviest CP-even scalar (e) and logarithm of the branching ratio of $\mu\rightarrow e \gamma$ (f). For all panels, the $x-$axis is displayed in logarithmic scale. All points shown here are also consistent with neutrino physics constraints as well as with constraints from the triple Higgs coupling measurements \cite{ATLAS-CONF-2022-050,ATLAS-CONF-2021-052}.}
		\label{fig:GW_plots}
	\end{figure*}

	First, we show in Fig.~\ref{fig:GW_plots} the spectral GW peak signal, $h^2\Omega_{\mathrm{GW}}^{\mathrm{peak}}$, as a function of the peak frequency, $f_{\mathrm{peak}}$ in Hz; both axes are shown in logarithmic scale. Focusing first on the panels (a), (b) and (c) where we display in the colour axis the relevant GW observables for the dynamics of the signal, namely, the phase transition strength (a), the inverse time-scale of the phase transition (b), and the nucleation temperature (c). Let us note that there are several planned GW interferometers that could explore the PGW signals predicted by the model. They include the Laser Interferometer Space Antenna (LISA) \cite{2017arXiv170200786A,Caprini:2019egz,Caprini:2015zlo}, the Big Bang Observer (BBO) \cite{Corbin:2005ny} and at the Deci-hertz Interferometer Gravitational wave Observatory (DECIGO) \cite{Kudoh:2005as}, indicating that the gravitational channel may offer a complementary approach to standard colliders channels in probing our model.  
	
\begin{figure*}[htb!]
		\begin{center}
			\includegraphics[width=0.7\textwidth]{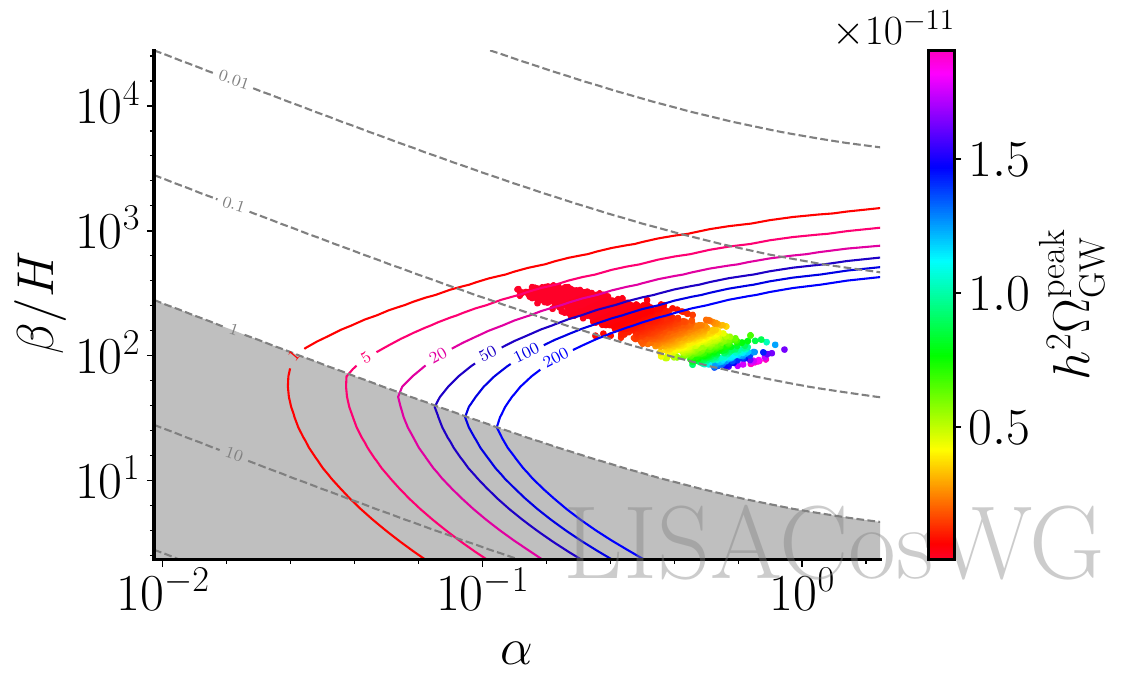}
		\end{center}
		\caption{Signal-to-noise ratio (SNR) plot for a LISA mission profile of 3 years. In the $x$-axis we show the strength of the phase transition ($\alpha$) and in the $y$-axis the inverse time duration in Hubble units. In the colour axis, we indicate the magnitude of the phase transition $h^2 \Omega^{\mathrm{peak}}_{\mathrm{GW}}$. The coloured isolines (corresponding to the values 1, 5, 20, 50, 100 and 200) are representative of the expected values for the SNR. As a general rule of thumb, points with an SNR greater than 5 can be potentially observed. The grey area corresponds to the region where the sound-wave component is dominant. All points shown in the plot correspond to the points potentially detectable by the LISA experiment (that is, above the red LISA curves of Fig.~\ref{fig:GW_plots}).}
		\label{fig:LISA_SNR}
	\end{figure*}

	As expected, a strong correlation between the GW observables and the signal strength is found. Indeed, higher values of $\alpha$ imply a stronger FOPT which in turn implies lower values for both the nucleation temperature $T_n$ and $\beta/H$. In particular, points which can be observed at LISA fall within the ranges $\alpha = [0.13,\, 0.88]$, $\beta/H = [81.13,\, 362.82]$ and $T_n = [99.77,\, 520.70]~\mathrm{GeV}$. For the points that fall within the LISA sensitivity domain, we have also computed the signal-to-noise ratio (SNR), as showcased in Fig.~\ref{fig:LISA_SNR}, where we note that for the strongest transitions that we found, the corresponding SNR is greater than 200, indicating that most of these points can indeed be probed by the LISA experiment \cite{weir_david_james_2022_6949107,Caprini:2019egz,Hindmarsh:2017gnf} 
	
	Now, focusing our attention on the bottom panels of Fig.~\ref{fig:GW_plots}, where we display various models parameters in the colour axis, namely, the VEV of the singlet field $\sigma$ (d), the mass of the second heaviest CP-even scalar $h_2$ (e) and the branching ratio $\mu\rightarrow e\gamma$ (f). Here, we note that there is no strong correlation between the physical parameters of the model and the GW observables. Nevertheless, we find that the higher values of the singlet VEV and of the mass of $h_2$, while allowing for the existence of FOPTs, give rise to the PGWs in higher frequency domains which fall outside the sensitivity of the future GW experiments. Indeed, for the parameter space regions which are accessible at LISA, the $\sigma$ VEV is typically below $5~\mathrm{TeV}$ whereas the $h_2$ mass is below $3~\mathrm{TeV}$. As for the last panel (f), we do not find any correlation with $\mu\rightarrow e\gamma$ process. Indeed, we find that various distinct possibilities for its branching ratio can be accommodated across different values of the frequency and the signal strength.
	
	In general, finding certain complementarity between the physical parameters of the model and PGW observables is relevant for exploring the model's parameter space. In the context of PGWs studies, it has been noted in Refs.~\cite{Zhou:2018zli,Ahriche:2018rao,Bernon:2017jgv,Biekotter:2022kgf} that potentially large corrections to the trilinear Higgs coupling compared to the SM expectation can significantly affect the type and the strength of the cosmological phase transitions. Indeed, strong FOPTs are typically driven by large trilinear scalar couplings as they strongly influence the potential barrier between different vacua.
	\begin{figure*}[htb!]
		\centering
		\begin{subfigure}[t]{0.48\textwidth}
			\centering
			\includegraphics[width=\textwidth]{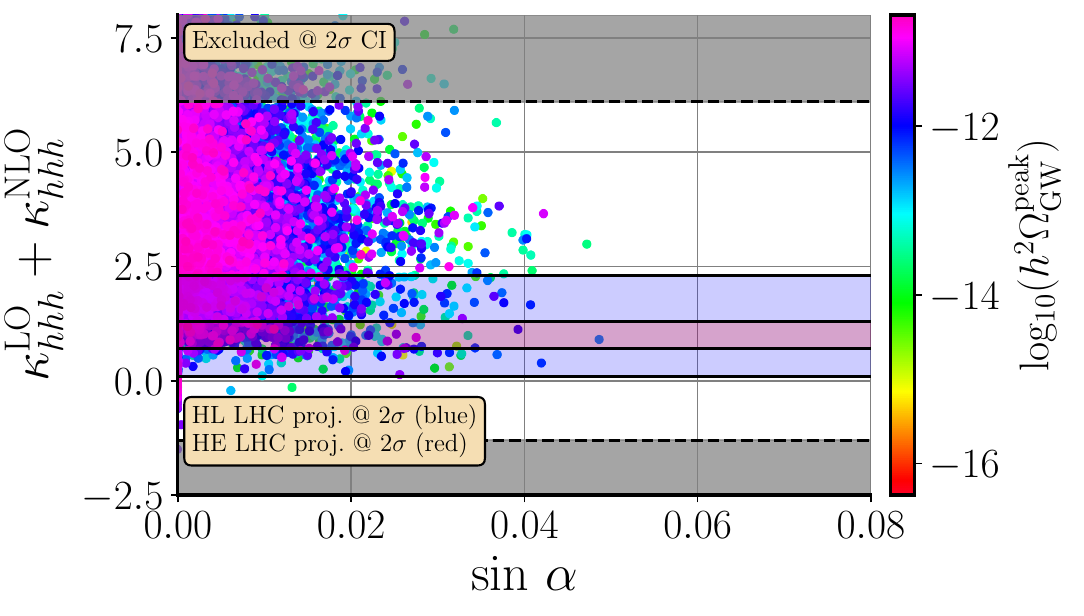}
			\caption{}
		\end{subfigure}
		\hfill
		\begin{subfigure}[t]{0.48\textwidth}
			\centering
			\includegraphics[width=\textwidth]{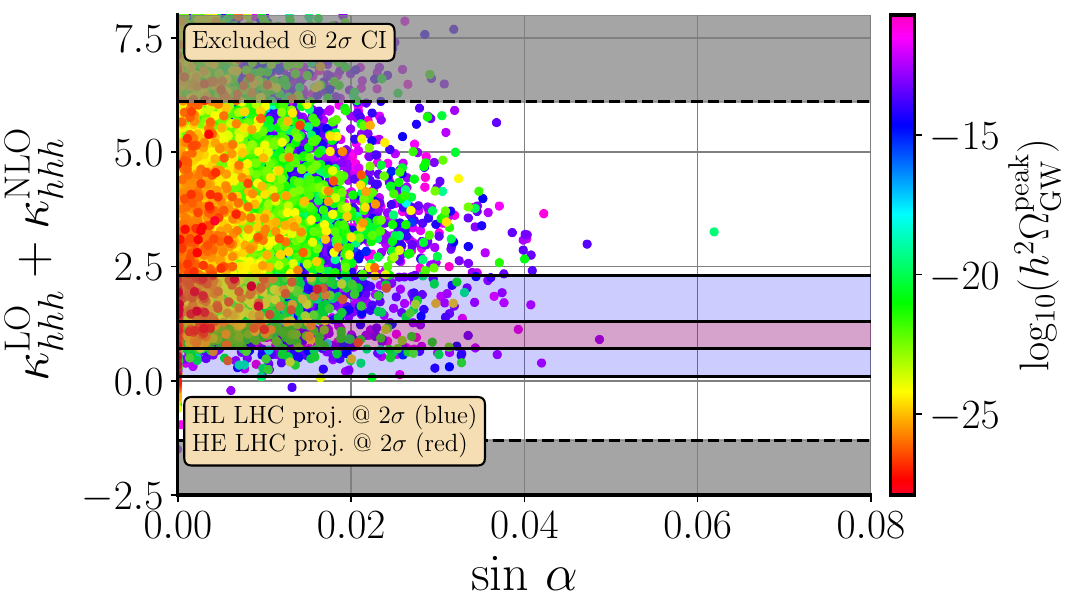}
			\caption{}
		\end{subfigure} 
		\caption{The BSM-to-SM ratio of trilinear Higgs couplings $\kappa \equiv \lambda^{\mathrm{BSM}}_{hhh}/\lambda^{\mathrm{SM}}_{hhh}$ as a function of the sine of the CP-even mixing angle $\alpha_h$. On the colour axis we show the logarithm of the GW intensity. On the panel (a), the results from the full collected dataset are showcased, whereas on (b) a cut on $h^2\Omega^{\mathrm{peak}}_{\mathrm{peak}}$ was imposed, such that only points which are visible at future GW experiments (LISA, BBO and DECIGO) are displayed. We note that all constraints from neutrino physics and LFV are applied here. Regions marked by gray bands are excluded at 95\% confidence level (CL) based on the latest results from the LHC, the blue band region corresponds to the expected sensitivity at the high-luminosity phase of the LHC \cite{ATL-PHYS-PUB-2022-018} while the red band region corresponds to the expected sensitivity at the high-energy phase of the LHC \cite{Homiller:2018dgu}.}
		\label{fig:Higgs_trilinear}
	\end{figure*}
	
	
	We show in Fig.~\ref{fig:Higgs_trilinear} the scatter plots for the BSM-to-SM ratio of the trilinear Higgs coupling computed for our BSM scenario versus its SM value, i.e.~$\kappa \equiv \lambda^{\mathrm{BSM}}_{hhh}/\lambda^{\mathrm{SM}}_{hhh}$, as a function of the sine of the CP-even mixing angle, with the logarithm of the GW signal strength in colour. Here, $\lambda_{hhh}^{\mathrm{BSM}} = \lambda_{hhh}^\mathrm{LO} + \lambda_{hhh}^\mathrm{NLO}$ is the full NLO result for the trilinear Higgs coupling in our model and $\lambda_{hhh}^{\mathrm{SM}}$ corresponds to the SM prediction. 
	
	Most of the generated points satisfy $\sin\alpha_h < 0.07$. Indeed, for such low values of the scalar mixing angle, Eq.~\eqref{eq:trilinear_higgs} asymptotically tends towards the SM prediction. However, NLO contributions result in a sizeable contribution to $\lambda_{hhh}$. We can also observe from panel (b) that the points with larger GW signal strength tend to accumulate as one approaches the alignment limit. However, this does not represent a solid trend as we find signal strengths well below the sensitivity ranges of LISA, BBO and DECIGO also for low values of the CP-even mixing angle, as can be readily noted in panel (a). Besides, one observes that there is a part of the parameter space that can already be excluded by the current data (gray bands) and it can be further reduced by future measurements both at the high-luminosity and high-energy phases of the LHC. However, the message to retain is that even within those regions there still exist parameter space points that feature FOPTs potentially detectable at future GW facilities.
	\begin{figure*}[htb!]
		\centering
		\begin{subfigure}[t]{0.31\textwidth}
			\centering
			\includegraphics[width=\textwidth]{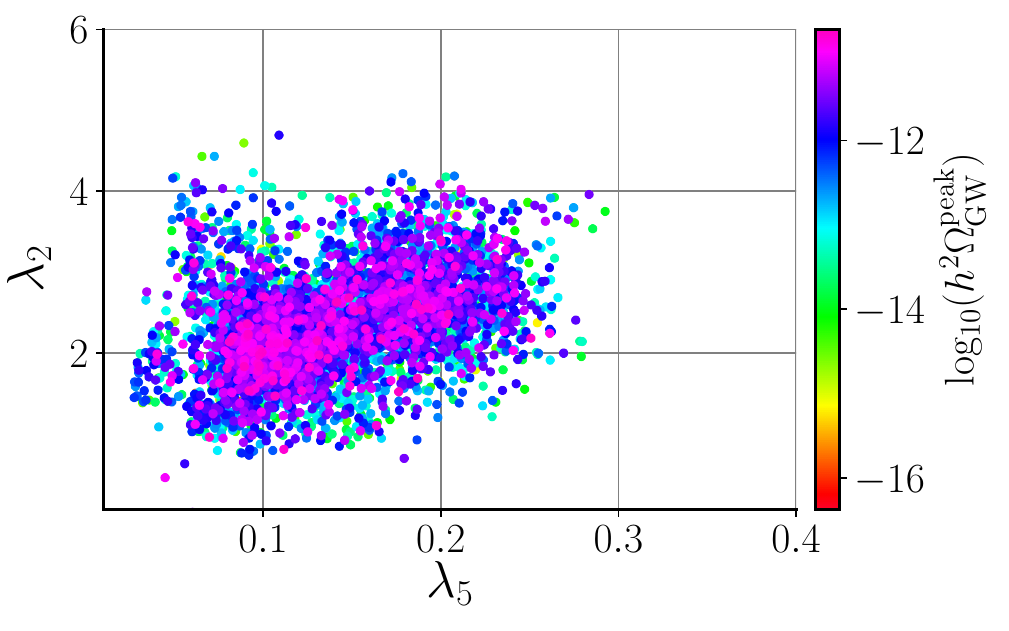}
			\caption{}
		\end{subfigure}
		\hfill
		\begin{subfigure}[t]{0.31\textwidth}
			\centering
			\includegraphics[width=\textwidth]{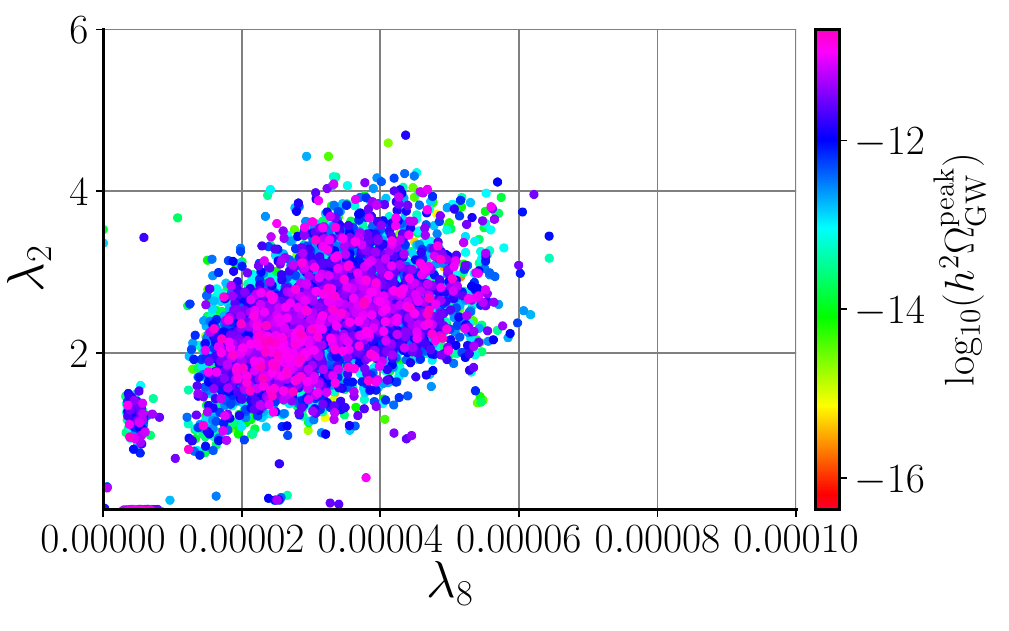}
			\caption{}
		\end{subfigure} 
		\hfill   
		\begin{subfigure}[t]{0.31\textwidth}
			\centering
			\includegraphics[width=\textwidth]{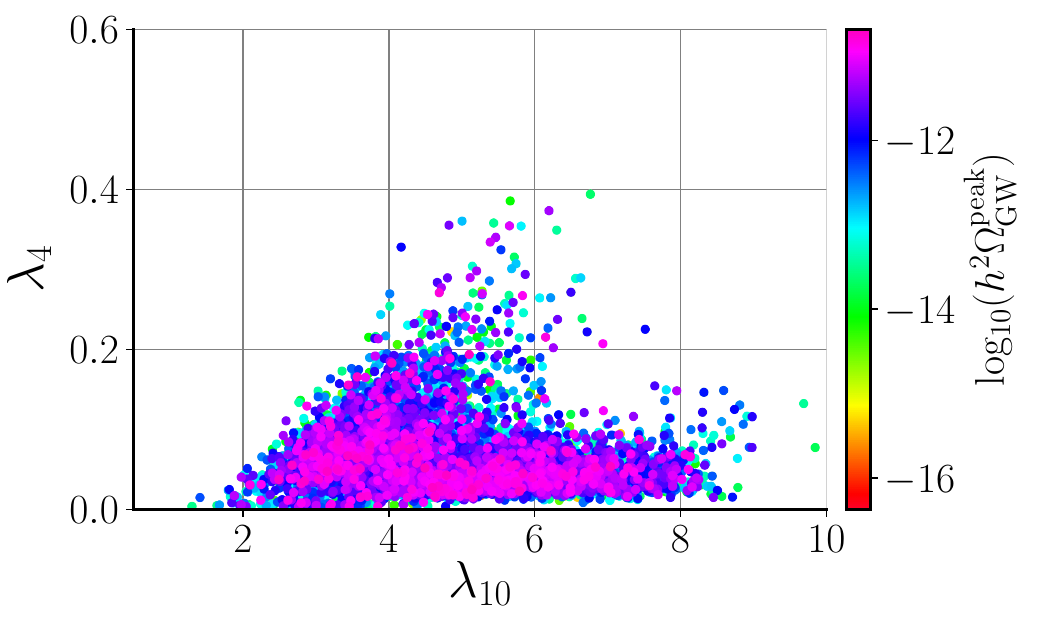}
			\caption{}
		\end{subfigure}
		\begin{subfigure}[t]{0.31\textwidth}
			\centering
			\hspace*{-1em}  
			\includegraphics[width=1.05\textwidth]{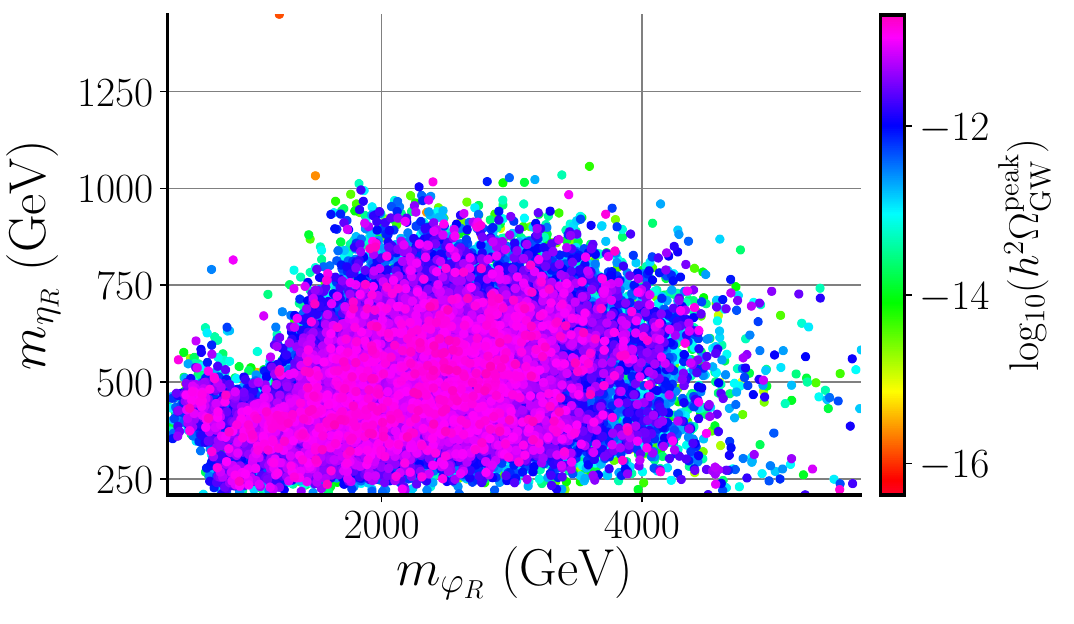}
			\caption{}
		\end{subfigure}
		\hfill
		\begin{subfigure}[t]{0.31\textwidth}
			\centering
			\includegraphics[width=\textwidth]{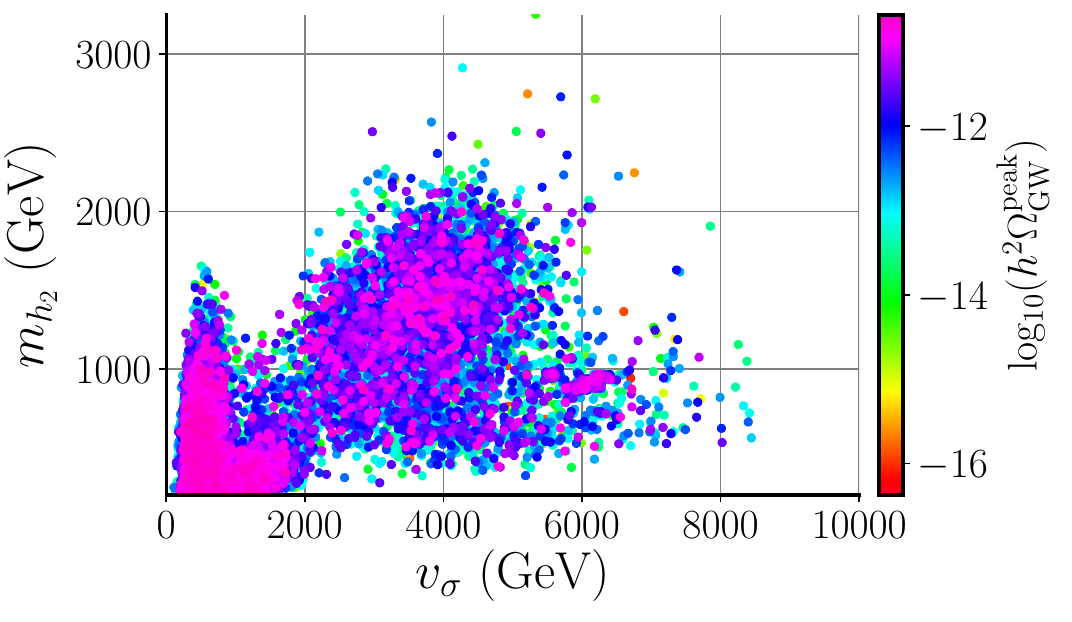}
			\caption{}
		\end{subfigure}
		\hfill
		\begin{subfigure}[t]{0.31\textwidth}
			\centering
			\includegraphics[width=\textwidth]{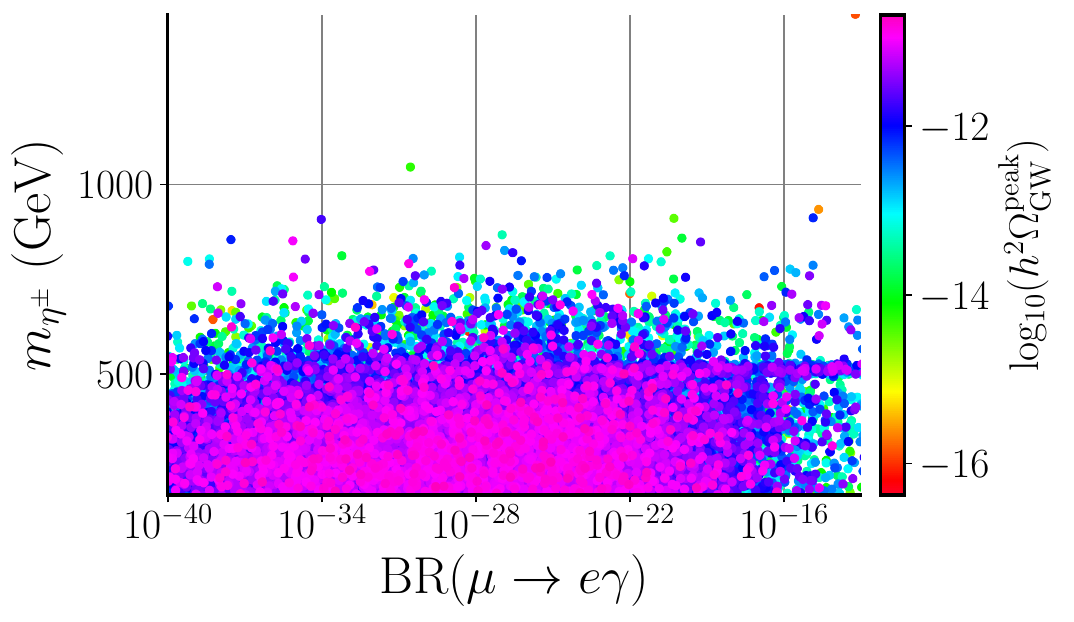}
			\caption{}
		\end{subfigure}
		\caption{Scatter plots of various physical parameters of the considered model (namely, quartic couplings, physical scalar masses and VEVs) with the logarithm of the GW intensity shown in the colour axis. In particular, we show $\lambda_2$ as a function of $\lambda_5$ (a), $\lambda_2$ as a function of $\lambda_8$ (b), $\lambda_4$ as a function of $\lambda_{10}$ (c), the mass of $\eta_R$ as a function of the $\varphi_R$ mass (d), the $h_2$ mass as a function of the singlet $\sigma$ VEV (e), and the mass of the charged scalar $\eta^\pm$ as a function of the branching ratio of $\mu \rightarrow e\gamma$ decay. The masses are given in units of GeV. Here, a cut on $h^2\Omega^{\mathrm{peak}}_{\mathrm{peak}}$ was imposed, such that only points which are visible at future GW experiments (either at LISA, BBO or DECIGO) are displayed. In panel (f), the $x$-axis is shown in logarithmic scale.}
		\label{fig:Lambdas_and_masses}
	\end{figure*}
	
	In Fig.~\ref{fig:Lambdas_and_masses} we show a selection of plots showing the impact of different quartic couplings and physical masses in the energy density amplitude of the GW spectra (colour axis). For panels (a), (b) and (c), we find various islands with distinct ranges of quartic couplings which lead to a strong PGW signal having both $\mathcal{O}(1)$ couplings, such as e.g.~$\lambda_2$ and $\lambda_{10}$ and much smaller couplings such as $\lambda_8$. Besides, we note that $\lambda_8$ is correlated with the size of the mixing for the CP-even states, and its smallness is related to the fact that we work very close to the alignment limit. Additionally, taking into account the results shown in Fig.~\ref{DMfig}, the constraints on $\lambda_8$ imposed by DM direct detection (both DARWIN and XENON1T) do not impact the presence of phase transitions and the corresponding GW spectra, since much lower values of $\lambda_8$ allow for the generation of phase transitions. Much tighter limits on the Higgs portal coupling would be needed if one would need to confront both DM and GW physics. In panels (d), (e) and (f) we notice that a wide range of physical scalar masses can be successfully accommodated with the observable GW signals. Let us stress that we did not find neither correlations nor further restrictions on the quartic couplings $\lambda_7$ and $\lambda_{13}$ involved in the neutrino mass generation. These parameters can take either small or large values without modifying the conclusions of this section. It can also be noted that no apparent correlation between the $\mu\rightarrow e\gamma$ decay rate and the mass of the charged scalar field is present, exhibiting a distinct flat distribution where most of the charged scalar masses tend to be below 1~TeV.
	
	\section{Conclusions}\label{sec:conclusions}
	
	In this work, we have presented a new scotogenic two-loop neutrino mass model where the SM gauge group is extended with an additional global $\mathrm{U(1)} \times \mathcal{Z}_2$ symmetry. The scalar sector of this framework contains the standard Higgs doublet, inert scalar doublet, singlet scalar fields as well as an active scalar singlet. Additionally, the model features the sector of right-handed Majorana neutrinos which in combination with the new exotic scalars leads to the generation of neutrino mass and mixing at two-loop level. The presence of Majorana neutrinos, together with a physical charged scalar field, also generates highly suppressed contributions to the tightly constrained LFV observables.
	
	The presence of such a rich scalar sector in this model opens the door for new possibilities for future new physics searches at collider, DM and GW experiments. In particular, our model features the generation of cosmological FOPTs in the early Universe that can lead to the formation of potentially observable stochastic PGWs. Focusing primarily on the phenomenologically favourable Higgs alignment limit, where the mixing between the CP-even states is small, we have performed a parameter scan of the model taking into account relevant constraints from LFV processes (such as $\mu\rightarrow e\gamma$ decay) and those from neutrino physics. We have explored the model's parameter space in detail and identified the domains that can successfully accommodate the existing constraints and giving rise to strong FOPTs that lead to observable PGWs at LISA. In addition, we have taken into account the constraints on the trilinear Higgs coupling at the NLO, and found a number of points with strong PGW signals that can be probed in both the collider and GW measurements in the future. We have found that the current constraints from the Higgs portal coupling does not impose severe constraints to the presence of observable GWs coming from phase transitions.
	
	\section*{Acknowledgments}
	\noindent
	A.E.C.H. acknowledges support by FONDECYT (Chile) under grant No.~1210378, Milenio-ANID-ICN2019\_044 and ANID PIA/APOYO AFB230003.
	The work of C.B. was supported by FONDECYT grant No. 1241855. 
	V.K.N. is supported by ANID-Chile Fondecyt Postdoctoral grant No. 3220005.
	R.P.~is supported in part by the Swedish Research Council grant, contract number 2016-05996, as well as by the European Research Council (ERC) under the European Union's Horizon 2020 research and innovation programme (grant agreement No 668679). 
	J.G. and A.P.M. are supported by the Center for Research and Development in Mathematics and Applications (CIDMA) through the Portuguese Foundation for Science and Technology (FCT - Funda\c{c}\~{a}o para a Ci\^{e}ncia e a Tecnologia), references UIDB/04106/2020 (\url{https://doi.org/10.54499/UIDB/04106/2020}) and UIDP/04106/2020 (\url{https://doi.org/10.54499/UIDP/04106/2020}). A.P.M. and J.G. are supported by the projects CERN/FIS-PAR/0021/2021 (\url{https://doi.org/10.54499/CERN/FIS-PAR/0021/2021}), CERN/FIS-PAR/0019/2021 (\url{https://doi.org/10.54499/CERN/FIS-PAR/0019/2021}) and CERN/FIS-PAR/0025/2021 (\url{https://doi.org/10.54499/CERN/FIS-PAR/0025/2021}).
	J.G. is also directly funded by FCT through the doctoral program grant with the reference 2021.04527.BD (\url{ttps://doi.org/10.54499/2021.04527.BD}).
	A.P.M.~is also supported by national funds (OE), through FCT, I.P., in the scope of the framework contract foreseen in the numbers 4, 5 and 6 of the article 23, of the Decree-Law 57/2016, of August 29, changed by Law 57/2017, of July 19 (\url{https://doi.org/10.54499/10.54499/DL57/2016/CP1482/CT001}). The authors also would like to acknowledge the FCT Advanced Computing Project to provide computational resources via the project CPCA/A00/7395/2020. This work was partially produced with the support of INCD funded by FCT and FEDER under the project 01/SAICT/2016 nº 022153. 
	C.B. would like to acknowledge the hospitality and support
	from the ICTP through the Associates Programme (2023-2028).
	
	\appendix
	
	\section{One-loop expression for the physical trilinear Higgs coupling}\label{app:one-loop}
	
	In the limit where the external momentum goes to zero, only the fields with mass near or above the Higgs field provide with leading corrections to the Higgs triple coupling, which in our model includes the top quark and all BSM scalar fields at the NLO. The neutral heavy Majorana fields do not contribute to the amplitude, since they do not couple directly to the Higgs boson before and after the EW symmetry breaking. Following Ref.~\cite{Camargo-Molina:2016moz}, the one-loop corrections involving the top quark reads as 
	\begin{equation}\label{eq:top_quark_1loop}
		\begin{aligned}
			\lambda_{hhh}^{t} = \frac{1}{32 \pi^2}\qty[
			-36 (\mathcal{F}(m_t, m_t)-1) \lambda^{(0)}_{tth} \lambda^{(0)}_{tthh}-24 \mathcal{F}(m_t, m_t, m_t) (\lambda^{(0)}_{tth})^3] \,,
		\end{aligned}
	\end{equation}
	where $m_t \approx 172.76~\mathrm{GeV}$ is the top mass. The NLO contribution coming from the scalar fields reads
	\begin{equation}\label{eq:Scalar_1loop}
		\begin{aligned}
			\lambda_{hhh}^{\phi} = \frac{1}{32\pi^2} \Big[ &4\mathcal{F}(m_{\eta^\pm}, m_{\eta^\pm}, m_{\eta^\pm}) (\lambda^{(0)}_{h_1 \eta^\pm \eta^\pm})^3 + 2 \sum_{k=2}^8 \mathcal{F}(m_k,m_k,m_k) (\lambda^{(0)}_{h_1 k k})^3 + \\ &6\mathcal{F}(m_{h_2},m_{h_2},m_{h_1})(\lambda^{(0)}_{h_1 h_1 h_2})^2 \sum_{k=2}^{3} \lambda_{h_1 k k}^{(0)} + 2\mathcal{F}(m_{\eta_R}, m_{\eta_R}, m_{\eta_R}) (\lambda^{(0)}_{h_1 \eta_R \eta_R})^3 - \\
			&3\sum_{k=2}^{8}\qty(\mathcal{F}(m_k, m_k) -1) \lambda_{h_1kk}^{(0)} \lambda_{h_1h_1kk}^{(0)} + 6\qty(\mathcal{F}(m_{\eta^\pm}, m_{\eta^\pm}) -1) \lambda_{h_1 \eta^\pm \eta^\pm}^{(0)} \lambda_{h_1 h_1 \eta^\pm \eta^\pm}^{(0)} + \\
			&6\qty(\mathcal{F}(m_{h_1}, m_{h_2}) -1) \lambda_{h_1 h_1 h_2}^{(0)} \lambda_{h_1 h_1 h_2 h_2}^{(0)} + \Bigg[6\sum_{k=\eta_R,\varphi_R} \lambda_{h_1 \eta_R k}^{(0)} \qty(\lambda_{h_1 k \varphi_R}^{(0)})^2 \mathcal{F}(m_{\phi_R}, m_{\eta_R}, k) + \\ &\qty(\mathrm{R}\rightarrow \mathrm{I})\Bigg] + \Big[6\lambda_{h_1 \eta_R \varphi_R}^{(0)} \lambda_{h_1 h_1 \eta_R \varphi_R}^{(0)} \mathcal{F}(m_{\eta_R}, m_{\varphi_R}) + \qty(\mathrm{R} \rightarrow \mathrm{I})\Big]
			\Big]\,,
		\end{aligned}
	\end{equation}
	where for simplicity of presentation we have $k = 1\dots 8 \equiv \eta^\pm, h_1, h_2, \eta_R, \varphi_R, \sigma_I, \eta_I, \varphi_I$. The superscript ``(0)'' indicates tree-level accuracy. The loop function $\mathcal{F}$ is defined as \cite{Camargo-Molina:2016moz}
	\begin{equation}\label{eq:loop_functionF}
		\mathcal{F}(m_{a_1}\dots m_{a_N}) = \sum_{x=1}^N \frac{m_{a_x}^2 \log\left(\frac{m_{a_x}^2}{\mu^2}\right)}{\Pi_{y \neq x} \left( m_{a_x}^2 - m_{a_y}^2 \right)}\,,
	\end{equation}
	where $\mu$ is the renormalisation scale. Here, we have used the same definition for the scale as in the calculation of the Coleman-Weinberg potential, i.e.~the renormalisation scale is defined as $\mu = \prod_{i=1}^{n} m_{\phi_i}^{1/n}$, where $i$ runs over all BSM scalars in the model. The various $\lambda$ couplings in the previous expressions are written as
	\begin{align}\label{eq:lambdas}
		&\lambda_{thh}^{(0)} = 4 \cos(\alpha_h) y_t^2 v\,, \\
		&\lambda_{tthh}^{(0)} = 4 \cos^2(\alpha_h) y_t^2\,, \\
		&\lambda_{h_1\eta^\pm \eta^\mp}^{(0)} = -\sin(\alpha_h)\lambda_{10}v_{\sigma} + \cos(\alpha_h)\lambda_5 v\,,\\
		&\lambda_{h_1h_1h_1}^{(0)} = 6\cos^3(\alpha_h) v \lambda_1 + 6\sin^3(\alpha_h) v_\sigma\lambda_4 + 3 \cos(\alpha_h) \sin^2(\alpha_h) v \lambda_8 + 3\cos^2(\alpha_h) \sin(\alpha_h) v_\sigma \lambda_8\,, \\
		\begin{split}
			&\lambda_{h_1h_2h_2}^{(0)} = 6 \text{$\lambda_1$} v \sin ^2(\alpha_h ) \cos (\alpha_h )+\text{$\lambda_8$} v \cos ^3(\alpha_h )-2 \text{$\lambda_8$} v \sin ^2(\alpha_h ) \cos (\alpha_h ) -\\ &\hspace{4.5em} -6 \text{$\lambda_4$} v_\sigma \sin (\alpha_h ) \cos ^2(\alpha_h )- \text{$\lambda_8$} v_\sigma  \sin ^3(\alpha_h )+ 2 \text{$\lambda_8$}  \sin (\alpha_h ) \cos ^2(\alpha_h )\,,
		\end{split}
		\\
		\begin{split}
			&\lambda_{h_1\eta_R\eta_R}^{(0)} = \cos ^2(\text{$\alpha_H$}) \qty[v \cos (\alpha_h) (\text{$\lambda_5$}+\text{$\lambda_7$})-\text{$\lambda_{10}$} \text{$v_\sigma$} \sin(\alpha_h)] + \\ &\hspace{4.5em} \sin^2(\text{$\alpha_H$}) \qty[\text{$\lambda_9$} v \cos(\alpha_h)-\text{$v_\sigma$} \sin (\alpha_h) (\text{$\lambda_{12}$}+\text{$\lambda_{13}$})] - \\ &\hspace{4.5em} \frac{1}{4} \sin(2\text{$\alpha_H$})(\text{$\lambda_{14}$}+\text{$\lambda_{15}$}) \qty[\text{$v_\sigma$} \cos (\alpha_h)-v \sin (\alpha )]\,,
		\end{split} \\
		\begin{split}
			&\lambda_{h_1\varphi_R\varphi_R}^{(0)} = \sin ^2(\text{$\alpha_H$}) \qty[v \cos (\alpha_h) (\text{$\lambda_5$}+\text{$\lambda_7$})-\text{$\lambda_{10}$} \text{$v_\sigma$} \sin(\alpha_h)] + \\ &\hspace{4.5em} \cos^2(\text{$\alpha_H$}) \qty[\text{$\lambda_9$} v \cos(\alpha_h)-\text{$v_\sigma$} \sin (\alpha_h) (\text{$\lambda_{12}$}+\text{$\lambda_{13}$})] + \\ &\hspace{4.5em} \frac{1}{4} \sin(2\text{$\alpha_H$})(\text{$\lambda_{14}$}+\text{$\lambda_{15}$}) \qty[\text{$v_\sigma$} \cos (\alpha_h)-v \sin (\alpha )]\,,
		\end{split} \\
		\begin{split}
			& \lambda_{h_1 \eta_R \varphi_R}^{(0)} = \frac{1}{4} \Big[2 \sin (2 \text{$\alpha_H$}) \qty{v \cos (\alpha_h)(\text{$\lambda_5$}+\text{$\lambda_7$}-\text{$\lambda_9$})+\text{$v_\sigma$} \sin (\alpha_h) (-\text{$\lambda_{10}$}+\text{$\lambda_{12}$}+\text{$\lambda_{13}$})} + \\ &\hspace{4.5em} \cos (2 \text{$\alpha_H$}) (\text{$\lambda_{14}$}+\text{$\lambda_{15}$}) \qty{\text{$v_\sigma$} \cos (\alpha_h)-v \sin (\alpha_h)}  \Big]\,,	
		\end{split} \\
		\begin{split}
			& \lambda_{h_1 \sigma_I \varphi_I}^{(0)} = \frac{1}{4} \Big[2 \sin (2 \text{$\beta_H$}) \qty{v \cos (\alpha_h)(\text{$\lambda_5$}-\text{$\lambda_7$}-\text{$\lambda_9$})-\text{$v_\sigma$} \sin (\alpha_h) (\text{$\lambda_{10}$}-\text{$\lambda_{12}$}+\text{$\lambda_{13}$})} - \\ &\hspace{4.5em} \cos (2 \text{$\beta_H$}) (\text{$\lambda_{14}$}-\text{$\lambda_{15}$})\qty{\text{$v_\sigma$} \cos (\alpha_h)-v \sin (\alpha_h) }\Big] \,,
		\end{split} \\
		\begin{split}
			&\lambda_{h_1\sigma_I\sigma_I}^{(0)} = \cos ^2(\text{$\beta_H$}) \qty[v \cos (\alpha_h) (\text{$\lambda_5$}-\text{$\lambda_7$})-\text{$\lambda_{10}$} \text{$v_\sigma$} \sin(\alpha_h)] +\\ &\hspace{4.5em} \sin^2(\text{$\beta_H$}) \qty[\text{$\lambda_9$} v \cos(\alpha_h)+\text{$v_\sigma$} \sin (\alpha_h) (\text{$\lambda_{13}$}-\text{$\lambda_{12}$})] + \\ &\hspace{4.5em} \frac{1}{4} \sin (2\text{$\beta_H$})(\text{$\lambda_{14}$}-\text{$\lambda_{15}$}) \qty[\text{$v_\sigma$} \cos (\alpha_h)-v \sin (\alpha_h)]\,,
		\end{split} \\
		&\lambda_{h_1\eta_I\eta_I}^{(0)} = v_1 \cos (\alpha_h ) (\text{$\lambda_5$}-\text{$\lambda_7$})-\text{$\lambda_{10}$} v_\sigma \sin (\alpha_h )\,, \\
		\begin{split}
			&\lambda_{h_1\varphi_I\varphi_I}^{(0)} = \frac{1}{4} v \sin (\alpha_h) \sin (2\text{$\beta_H$}) (\text{$\lambda_{14}$}-\text{$\lambda_{15}$})+\sin ^2(\text{$\beta_H$}) \qty[v \cos(\alpha_h) (\text{$\lambda_5$}-\text{$\lambda_7$})-\text{$\lambda_{10}$}\text{$v_\sigma $} \sin (\alpha_h)]+ \\ &\hspace{4.5em} \cos^2(\text{$\beta_H$}) \qty[\text{$\lambda_9$} v\cos(\alpha_h)+\text{$v_\sigma$} \sin(\alpha_h)(\text{$\lambda_{13}$}-\text{$\lambda_{12}$})] + \\ &\hspace{4.5em} \frac{1}{2} \text{$v_\sigma$}\cos(\alpha_h) \sin (\text{$\beta_H$}) \cos(\text{$\beta_H$})(\text{$\lambda_{15}$}-\text{$\lambda_{14}$})\,, 
		\end{split}\\
		\begin{split}
			&\lambda_{h_1 h_1 h_2}^{(0)} = 6 \text{$\lambda_1$} v \sin (\alpha_h ) \cos ^2(\alpha_h )+\text{$\lambda_8$} v \sin ^3(\alpha_h )-2 \text{$\lambda_8$} v \sin (\alpha_h ) \cos ^2(\alpha_h )+ \\ &\hspace{4.5em} 6 \text{$\lambda_4$} v_\sigma \sin ^2(\alpha_h ) \cos (\alpha_h) + \text{$\lambda_8$} v_\sigma \cos ^3(\alpha_h )- 2 \text{$\lambda_8$} v_\sigma \sin ^2(\alpha_h ) \cos (\alpha_h )\,,
		\end{split}
		\\
		&\lambda_{h_1 h_1 \eta^\pm \eta^\pm}^{(0)} = \text{$\lambda_{10}$} \sin ^2(\alpha_h )+\text{$\lambda_5$} \cos ^2(\alpha_h )\,, \\
		&\lambda_{h_1 h_1 h_1 h_1}^{(0)} = 6 \text{$\lambda_1$} \cos ^4(\alpha_h )+6 \text{$\lambda_4$} \sin ^4(\alpha_h )+6 \text{$\lambda_8$} \sin ^2(\alpha_h ) \cos ^2(\alpha_h )\,, \\
		\begin{split}
			&\lambda_{h_1 h_1 h_2 h_2}^{(0)} = 6 \text{$\lambda_1$} \sin ^2(\alpha_h ) \cos ^2(\alpha_h )+6 \text{$\lambda_4$} \sin ^2(\alpha_h ) \cos ^2(\alpha_h )+\text{$\lambda_8$} \sin ^4(\alpha_h )+\text{$\lambda_8$} \cos ^4(\alpha_h )- \\ &\hspace{4.5em} 4 \text{$\lambda_8$} \sin ^2(\alpha_h ) \cos ^2(\alpha_h )\,,
		\end{split} \\
		\begin{split}
			&\lambda_{h_1 h_1 h_1 h_2}^{(0)} = \frac{3}{2} \sin (2 \alpha_h) \qty[\cos (2 \alpha_h) (\text{$\lambda_1$}+\text{$\lambda_4$}-\text{$\lambda_8$})+\text{$\lambda_1$}-\text{$\lambda_4$}]\,,
		\end{split} \\
		\begin{split}
			&\lambda_{h_1 h_1 \eta_R \eta_R}^{(0)} = \sin ^2(\alpha_h) \qty[\text{$\lambda_{10}$} \cos^2(\text{$\alpha_H$})+\sin^2(\text{$\alpha_H$}) (\text{$\lambda_{12}$}+\text{$\lambda_{13}$})] + \\ &\hspace{4.5em} \sin (\alpha_h) \cos (\alpha_h) \sin (\text{$\alpha_H$}) \cos(\text{$\alpha_H$}) (\text{$\lambda_{14}$}+\text{$\lambda_{15}$}) + \\ &\hspace{4.5em} \cos^2(\alpha_h) \qty[\cos^2(\text{$\alpha_H$}) (\text{$\lambda_5$}+\text{$\lambda_7$})+\text{$\lambda_9$} \sin^2(\text{$\alpha_H$})]\,,
		\end{split} \\
		\begin{split}
			&\lambda_{h_1 h_1 \varphi_R \varphi_R}^{(0)} = \sin ^2(\alpha_h) \qty[\text{$\lambda_{10}$} \sin^2(\text{$\alpha_H$})+\cos^2(\text{$\alpha_H$}) (\text{$\lambda_{12}$}+\text{$\lambda_{13}$})] - \\ &\hspace{4.5em} \sin (\alpha_h) \cos (\alpha_h) \sin (\text{$\alpha_H$}) \cos(\text{$\alpha_H$}) (\text{$\lambda_{14}$}+\text{$\lambda_{15}$}) + \\ &\hspace{4.5em} \cos^2(\alpha_h) \qty[\sin^2(\text{$\alpha_H$}) (\text{$\lambda_5$}+\text{$\lambda_7$})+\text{$\lambda_9$} \cos^2(\text{$\alpha_H$})]\,, 
		\end{split} \\
		\begin{split}	
			&\lambda_{h_1 h_1 \sigma_I \sigma_I}^{(0)} = \sin^2(\alpha_h) \qty[\text{$\lambda_{10}$} \cos^2(\text{$\beta_H$})+\sin^2(\text{$\beta_H$}) (\text{$\lambda_{12}$}-\text{$\lambda_{13}$})] + \\ &\hspace{4.5em} \sin(\alpha_h) \cos (\alpha_h) \sin (\text{$\beta_H$}) \cos (\text{$\beta_H$})(\text{$\lambda_{15}$}-\text{$\lambda_{14}$}) + \\&\hspace{4.5em} \cos^2(\alpha_h) \qty[\cos^2(\text{$\beta_H$}) (\text{$\lambda_5$}-\text{$\lambda_7$})+\text{$\lambda_9$} \sin^2(\text{$\beta_H$})]\,,
		\end{split} \\
		&\lambda_{h_1 h_1 \eta_I \eta_I}^{(0)} = \text{$\lambda_{10}$} \sin ^2(\alpha_h )+\cos ^2(\alpha_h ) (\text{$\lambda_5$}-\text{$\lambda_7$})\,, \\
		\begin{split}
			&\lambda_{h_1 h_1 \varphi_I \varphi_I}^{(0)} = \sin ^2(\alpha_h) \qty[\text{$\lambda_{10}$} \sin^2(\text{$\beta_H$})+\cos^2(\text{$\beta_H$}) (\text{$\lambda_{12}$}-\text{$\lambda_{13}$})] + \\ &\hspace{4.5em} \sin(\alpha_h) \cos (\alpha_h) \sin (\text{$\beta_H$}) \cos (\text{$\beta_H$})(\text{$\lambda_{14}$}-\text{$\lambda_{15}$}) + \\ &\hspace{4.5em} \cos^2(\alpha_h) \qty[\sin^2(\text{$\beta_H$}) (\text{$\lambda_5$}-\text{$\lambda_7$})+\text{$\lambda_9$} \cos^2(\text{$\beta_H$})]\,,
		\end{split} \\
		\begin{split}
			&\lambda_{h_1 h_1 \eta_R \varphi_R}^{(0)} = \sin (\text{$\alpha_H$}) \cos (\text{$\alpha_H$}) \qty[\sin ^2(\alpha_h)(\text{$\lambda_{10}$}-\text{$\lambda_{12}$}-\text{$\lambda_{13}$})+\cos^2(\alpha_h) (\text{$\lambda_5$}+\text{$\lambda_7$}-\text{$\lambda_9$})] - \\ &\hspace{4.5em} \frac{1}{4} \sin (2 \alpha_h) \cos (2 \text{$\alpha_H$})(\text{$\lambda_{14}$}+\text{$\lambda_{15}$})\,,	
		\end{split} \\
		\begin{split}
			&\lambda_{h_1 h_1 \sigma_I \varphi_I}^{(0)} = \frac{1}{4} (\sin (2 \alpha_h) \cos (2 \text{$\beta_H$}) (\text{$\lambda_{14}$}-\text{$\lambda_{15}$})- \\ &\hspace{4.5em} \sin (2 \text{$\beta_H$}) \qty[\cos (2 \alpha_h)(\text{$\lambda_{10}$}-\text{$\lambda_{12}$}+\text{$\lambda_{13}$}-\text{$\lambda_5$}+\text{$\lambda_7$}+\text{$\lambda_9$})-\text{$\lambda_{10}$}+\text{$\lambda_{12}$}-\text{$\lambda_{13}$}-\text{$\lambda_5$}+\text{$\lambda_7$}+\text{$\lambda_9$}]) \,,
		\end{split}
	\end{align}
	such that the trilinear Higgs coupling at one-loop accuracy is determined as $\lambda_{hhh}^{\mathrm{NLO}} = \lambda_{h_1h_1h_1}^{(0)} + \lambda_{hhh}^{t} + \lambda_{hhh}^{\phi}$. Here, $\alpha_H$ is the mixing angle between the heavy $\eta_R$ and $\varphi_R$ fields and $\beta_H$ is the mixing angle between the CP-odd fields $\eta_I$ and $\varphi_I$.

	\bibliographystyle{JHEP}
	\bibliography{IDMtwoloop}
	
\end{document}